\documentclass[aps,prd,reprint,superscriptaddress]{revtex4-1}

\usepackage{graphicx}
\usepackage{dcolumn}
\usepackage{natbib,aas_macros}
\usepackage{xcolor}
\usepackage{amssymb}
\usepackage{amsthm}
\usepackage{mathrsfs}
\usepackage{amsmath}
\usepackage{lineno}
\usepackage{hyperref}
\hypersetup{
    pdfnewwindow=true,      
    colorlinks=true,        
    linkcolor=blue,         
    citecolor=blue,        
    filecolor=blue,         
    urlcolor=blue           
}

\newcommand{\Hs}{H_{\rm s}}
\newcommand{\Ho}{H_{\rm 0}}
\newcommand{\Hc}{H_{\rm c}}
\newcommand{\Hcnu}{H_{\rm c, \nu}}
\newcommand{\Hcgw}{H_{\rm c, gw}}
\newcommand{\dir}{{\bf \Omega}}
\newcommand{\xgw}{{\bf x}_{\rm gw}}

\newcommand{\xnui}{{\bf x}_{\nu i}}
\newcommand{\xnuj}{{\bf x}_{\nu j}}
\newcommand{\Xnu}{{\bf X}_{\rm \nu}}

\newcommand{\teta}{{\boldsymbol \theta}}

\definecolor{comment}{RGB}{166, 38, 164}

\begin{document}
\title{Bayesian Multi-Messenger Search Method for Common Sources of \\ Gravitational Waves and High-Energy Neutrinos}

\author{Imre Bartos}
\thanks{Email: imrebartos@ufl.edu}
\address{Department of Physics, University of Florida, Gainesville, FL 32611}

\author{Do\u{g}a Veske}
\address{Department of Physics, Columbia University, New York, NY 10027}

\author{Azadeh Keivani}
\address{Department of Physics, Columbia University, New York, NY 10027}

\author{Zsuzsa M\'arka}
\address{Department of Physics, Columbia University, New York, NY 10027}

\author{Stefan Countryman}
\address{Department of Physics, Columbia University, New York, NY 10027} 

\author{Erik Blaufuss}
\address{Department of Physics, University of Maryland, College Park, MD 20742, USA}

\author{Chad Finley}
\address{Oskar Klein Centre and Department of Physics, Stockholm University, SE-10691 Stockholm, Sweden}

\author{Szabolcs M\'arka}
\address{Department of Physics, Columbia University, New York, NY 10027}

\begin{abstract}
Multi-messenger astrophysics is undergoing a transition towards low-latency searches based on signals that could not individually be established as discoveries. The rapid identification of signals is important in order to initiate timely follow-up observations of transient emission that is only detectable for short time periods. Joint searches for gravitational waves and high-energy neutrinos represent a prime motivation for this strategy. Both gravitational waves and high-energy neutrinos are typically emitted over a short time frame of seconds to minutes during the formation or evolution of compact objects. In addition, detectors searching for both messengers observe the whole sky continuously, making observational information on potential transient sources rapidly available to guide follow-up electromagnetic surveys. The direction of high-energy neutrinos can be reconstructed to sub-degree precision, making a joint detection much better localized than a typical gravitational wave signal. Here we present a search strategy for joint gravitational wave and high-energy neutrino events that allows the incorporation of astrophysical priors and detector characteristics following a Bayesian approach. We aim to determine whether a multi-messenger correlated signal is a real event, a chance coincidence of two background events or the chance coincidence of an astrophysical signal and a background event. We use an astrophysical prior that is model agnostic and takes into account mostly geometric factors. Our detector characterization in the search is mainly empirical, enabling detailed realistic accounting for the sensitivity of the detector that can depend on the source properties. By this means, we will calculate the false alarm rate for each multi-messenger event which is required for initiating electromagnetic follow-up campaigns. 
\end{abstract}
\maketitle

\section{Introduction}
Multi-messenger astrophysics produced two foundational discoveries in 2017: the detection of a binary-neutron star merger through gravitational waves (GWs) and electromagnetic emission \cite{2017ApJ...848L..12A}, and the observation of a blazar through high-energy neutrinos and electromagnetic emission \cite{ic1709022mm}. The multimessenger science reach of the GW detectors had been enabled by decades of effort preceding the discovery~\cite{LIGOG060660,2006ivoa.spec.1101S,2008CQGra..25k4039A,2008CQGra..25k4051A,2009IJMPD..18.1655V,2010JPhCS.243a2001M,2011APh....35....1B,2011CQGra..28k4013M,2011GReGr..43..437C,2011ivoa.spec.0711S,2011PhRvL.107y1101B,2012JPhCS.363a2022B,
2012PhRvD..85j3004B,2012PhRvD..86h3007B,2013APh....45...56S,2013CQGra..30l3001B,2013JCAP...06..008A,2013PhRvL.110x1101B,2013RvMP...85.1401A,2014PhRvD..90j1301B,2014PhRvD..90j2002A,2015PhRvL.115w1101B,2016PhRvD..93l2010A,2017ApJ...848L..12A,2017ApJ...850L..35A,2017PhRvD..96b2005A,2017PhRvD..96b3003B}.  

The third leg of multi-messenger astrophysics will be the discovery of GWs and high-energy neutrinos from a common source \cite{2013RvMP...85.1401A,2013CQGra..30l3001B}. Such a detection could shed light to, e.g., how newly formed compact objects accelerate particles to extreme energies. In addition, some high-energy neutrinos are identified rapidly with localization accuracies much better than that available with GW detectors, which can guide observatories in their search of the electromagnetic counterparts of GW sources.

Several source candidates are considered to generate GWs and high-energy neutrinos, including core-collapse supernovae~\cite{2012PhRvD..86h3007B, kohta-sn-2018}, gamma-ray bursts (GRBs)~(See e.g.~\cite{kohta2006,meszaros13}), BNS mergers~\cite{2018arXiv180511613K}, neutron star-black hole  mergers~\cite{shigeo2017}, soft gamma repeaters~\cite{ioka2005,murphy2013}, and microquasars~\cite{2011GReGr..43..437C}. 
Besides these candidate sources, searches might reveal unknown source populations or production mechanisms.
Detecting even one joint source of GWs and high-energy neutrinos will significantly increase our understanding of the underlying mechanisms that create them~\cite{2013RvMP...85.1401A,2013CQGra..30l3001B}.

Searching for joint GW+high-energy neutrino (hereafter GW+neutrino) sources has only become viable in recent years with the advent of large-scale detectors, in particular the Advanced LIGO \cite{aligo2015} and Advanced Virgo \cite{avirgo2015} observatories on the GW side, and the IceCube \cite{icecube2017}, ANTARES \cite{2011NIMPA.656...11A} and Pierre Auger \cite{2015arXiv150201323T} observatories on the neutrino side. Both sides will experience significant upgrades in the coming years. Advanced LIGO and Advanced Virgo are set to reach their design sensitivities within the next few years \cite{lrr2018}. IceCube started an upgrade towards a second generation detector, IceCube-Gen2, with several times improved sensitivity \cite{2014arXiv1412.5106I}. Another neutrino detector, KM3NeT, is being constructed in the Mediterranean \cite{2016JPhG...43h4001A}. Due to these advances, our reach to GW and neutrino sources is set to rapidly increase in the near future and beyond.

While no joint GW+neutrino discovery has been confirmed to date, there has been significant effort to search for such events. Following the first observational constraints on common sources in 2011 \cite{2011PhRvL.107y1101B}, independent searches were carried out using Initial LIGO/Virgo and the partially completed ANTARES and IceCube detectors \cite{2013JCAP...06..008A,2014PhRvD..90j2002A}. With the completion of Advanced LIGO, several searches were carried out to find the neutrino counterpart of GW discoveries \cite{2016PhRvD..93l2010A,2017PhRvD..96b2005A,2017ApJ...850L..35A}. A separate search was carried out to find joint events for which neither the GW nor the neutrino signal could be independently confirmed to be astrophysical \cite{Albert_2019}.

Most of these searches were based on the analysis method developed by Baret et al. \cite{2012PhRvD..85j3004B}. This method combines GW amplitude, neutrino reconstructed energy, temporal coincidence and directional coincidence to separate astrophysical events from chance coincidences. The method aims to be emission model agnostic and does not impose constraints on the source properties except by assuming that higher neutrino energy is more likely to indicate an astrophysical signal. 

Following the success of the search method by Baret et al. \cite{2012PhRvD..85j3004B} spanning over a decade, it is time to upgrade it to enhance its sensitivity and aid newly relevant real-time searches. Two particular motivations for the upgrade are to facilitate the incorporation of astrophysical information and detector characteristics in the search. Regarding astrophysical information, while it is beneficial to keep the search largely model independent, in many cases signal constraints can be specified that do not depend strongly on particular model. Regarding detector characteristics, a more complex detector model will improve sensitivity and accuracy, but requires the incorporation of prior information on these characteristics to the search.

In this paper, we present a new search algorithm for common sources of GWs and high-energy neutrinos based on Bayesian hypothesis testing. A Bayesian framework is a natural choice to incorporate prior astrophysical and detector information. Bayesian solutions are becoming more common in GW \cite{2010PhRvD..81f2003V,2005PhRvD..72j2002D,2015CQGra..32m5012C,2016PhRvD..93b4013S} and more recently multi-messenger data analysis \cite{2008ApJ...679..301B,2018ApJ...860....6A,1997scma.conf.....B,2013ApJS..209...30N,2014ApJ...795...43F,2017PhRvL.119r1102F}.

The paper is organized as follows. The general idea for this analysis is described in Sec~\ref{sec:bayes}, following by probabilities describing signal hypothesis in Sec~\ref{sec:sig}, null hypothesis in Sec~\ref{sec:null} and chance coincidence hypothesis in Sec~\ref{sec:coincidence}. We define the use of odds ratios in Section \ref{sec:odds}. We conclude in Section \ref{sec:conclusion}.

\section{Multi-messenger Search Method}
\label{sec:bayes}
To determine whether a multi-messenger coincident signal is a real event or a random coincidence, we formulate the problem in the context of Bayesian hypothesis testing. We further incorporate detector and background characteristics as well as astrophysical information of the messenger particle and its source. 

We will compare multiple hypotheses. Our signal hypothesis, $\Hs$, is that all considered messengers originated from the same astrophysical source. Our null hypothesis, $\Ho$, is that triggers in all messengers arose from the background. Additionally, we will consider a chance coincidence hypothesis, $\Hc$, that one type of the messengers has an astrophysical trigger, but the other type of messenger only has triggers from the background. We will neglect the possibility that different messengers from distinct astrophysical signals coincide as this is highly unlikely given our low signal rate.

For GWs we use the following observational information for the search: (i) detection time $t_{\rm gw}$; (ii) reconstructed sky location probability density $\mathcal{P}_{\rm gw}=\mathcal{P}_{\rm gw}(\dir)$, called the {\it skymap}, where $\dir$ is the source sky location; (iii) the GW data analysis pipeline specific signal-to-noise ratio (SNR) of the GW event in the GW detectors network $\rho_{\rm gw}$, which is the individual SNRs of the signals at each detector summed in quadrature, and (iv) reconstructed distance distribution $\mathcal{D}_{\rm gw}=\mathcal{D}_{\rm gw}(r)$ where \(r\) is the distance of the event to Earth \cite{PhysRevLett.116.061102}. We define a vector containing the measured properties of a GW trigger as
\begin{equation}
\xgw = \{t_{\rm gw}, \mathcal{P}_{\rm gw}, \rho_{\rm gw}, \mathcal{D}_{\rm gw}\}.
\end{equation}
For multiple source types, an additional variable could be the source-dependent gravitational waveform. We omit this as a factor in the following description.

For high-energy neutrinos, the used observational information for them includes (i) their detection times $t_{\nu}$; (ii) their reconstructed sky location probability densities $\mathcal{P}_{\nu}=\mathcal{P}_{\nu}(\dir)$; and (iii) their reconstructed neutrino energies $\epsilon_{\nu}$. As high-energy neutrinos are not directly observed, the observed energies of the leptons produced in the neutrino interactions are taken as $\epsilon_{\nu}$.  Generally the reconstructed neutrino sky location, $\mathcal{P}_{\nu}$, can be described as a Gaussian distribution centered at reconstructed neutrino direction $\dir_{\nu}$, with reconstructed uncertainty $\sigma_{\nu}$ \cite{2008APh....29..299B,2016JInst..1111009I}. We define a matrix containing the measured properties of all neutrino triggers as 
\begin{equation}
    \Xnu=\begin{bmatrix} {\bf x}_{\rm \nu 1} \\ {\bf x}_{\rm \nu 2} \\ ...\end{bmatrix}
\end{equation}
with rows
\begin{equation}
\xnui = \{t_{\rm \nu i}, \dir_{\rm \nu i}, \sigma_{\rm \nu i}, \epsilon_{\rm \nu i}\}.
\end{equation}
Throughout the paper we will assume we have \(N\) neutrino triggers.
We define a vector containing our model parameters for the signal hypothesis as
\begin{equation}
\teta=\{t_{\rm s}, r, \dir, E_{\rm gw}, E_{\rm \nu}\},
\end{equation}
where $t_{\rm s}$ is the reference time, $r$ is the luminosity distance, $\dir$ is the sky location, $E_{\rm gw}$ is the isotropic-equivalent total GW energy, and $E_{\nu}$ is the isotropic-equivalent total high-energy neutrino energy emitted from the astrophysical event. The reference time can be thought of as the time of a relevant astrophysical event to which we compare the other times of arrival, delayed by the travel time of information to Earth at the speed of light.
The neutrino energies considered here render the neutrino travel time practically the same as travel time at the speed of light.

At the end of our analysis we will compute a Bayes factor for our signal hypothesis given the observational data as 
\begin{equation}
\label{eq:or}
\mathcal{O_{\rm gw+\nu}} = \frac{P(\Hs|\xgw, \Xnu)}{P(\Ho|\xgw, \Xnu) + P(\Hc|\xgw, \Xnu)}.
\end{equation}

\section{Signal hypothesis}
\label{sec:sig}

We first introduce our signal hypothesis $\Hs$. This hypothesis considers having at least one coincident signal neutrino with the gravitational wave which is also signal. Therefore we split this hypothesis into sub-hypotheses for different number of coincident signal neutrinos and denote them by $\Hs^{n}$ where n is the number of coincident neutrinos. What this means is for example for n=1 we would have one neutrino which comes from the same source of gravitational wave and other neutrinos belong to background, or to the null hypothesis. In order to label the signal neutrinos and the background neutrinos separately we will use the notation $\Xnu^{\rm i}=\Xnu\setminus\xnui$ to refer to the $\Xnu$ matrix without the $\mbox{i}^{\rm th}$ row $=\xnui$. Given the observational data, the probability of the signal hypothesis being true can be written as 
\begin{equation}P(\Hs|\xgw, \Xnu)=\sum_{n=1}^{N} P(\Hs^{n}|\xgw, \Xnu)
\end{equation}
We use Bayes' rule to this expression
\begin{equation}
\sum_{n=1}^{N} P(\Hs^{n}|\xgw, \Xnu)=\sum_{n=1}^{N} \frac{P(\xgw,\Xnu|\Hs^{n})P(\Hs^{n})}{P(\xgw,\Xnu)}
\label{eq:1}
\end{equation}
We are interested in the ratio of such probabilities for different hypotheses, hence the denominator above will cancel out. We therefore omit its computation. 
Then we further expand the first term by specifying the signal neutrinos
\begin{widetext}
\begin{equation}
P(\xgw,\Xnu|\Hs^{n})=\sum_{\{i,j,...\}}P(\xgw,\Xnu|\Hs^{n},s=\{i,j,...\})P(s=\{i,j,...\}|\Hs^{n})
\label{eq:1.5}
\end{equation}
\end{widetext}
The sum in Eq. \ref{eq:1.5} is over all \({N \choose n}\) subsets of the set of integers from 1 to \(N\) with \(n\) elements, which are denoted with the set \(s\) in the sum which stands for the signal set
The second term on the right hand side corresponds to probability of each combination which is
\begin{equation}
P(s=\{i,j,...\}|\Hs^{n})={N \choose n}^{-1}
\label{eq:1.55}
\end{equation}
We further decompose the first term in Eq. \ref{eq:1.5} by separating the signal and background neutrino terms via using their independence with a memoryless detector assumption as
\begin{equation}
\begin{split}
P(\xgw,\Xnu|\Hs^{n},s=\{i,j,...\})=\\P(\xgw,\xnui,\xnuj,...|\Hs^{n})P(\Xnu^{i,j,...}|\Ho)
\end{split}
\label{eq:1.6}
\end{equation}
In Eq. \ref{eq:1.6} we dropped the \(s\) set from the conditions of the first term on the right side of the equation since there are already only \(n\) neutrinos in that probability, for convenience. Second term corresponds to the probability for all other than those \(n\) neutrinos belonging to background. Next, to obtain the first term in the right side of Eq. \ref{eq:1.6} we marginalize over the parameters
\begin{equation}
\begin{split}
P(\xgw, \xnui,\xnuj...|\Hs^{n}) = \\ \int P(\xgw, \xnui,\xnuj...|\teta, \Hs^n)P(\teta|\Hs^n)d\teta
\label{eq:2}
\end{split}
\end{equation}
Since $\xgw$ and all $\xnui$ which belong to signal hypothesis are dependent on $\teta$ but are otherwise can be considered independent of each other, we can separate the GW and the high-energy parts from each other such as
\begin{equation}
\begin{split}
P(\xgw, \xnui,\xnuj...|\teta, \Hs^n) = \\ P(\xgw|\teta, \Hs^n)P(\xnui|\teta, \Hs^n)P(\xnuj|\teta, \Hs^n)...
\label{eq:3}
\end{split}
\end{equation}
We now specify the independent elements of Eqs. \ref{eq:1}, \ref{eq:2} and \ref{eq:3} in the context of our astrophysical and detection models. \( P(\Xnu^{i,j,...}|\Ho)\) in Eq. \ref{eq:1.6} will be explained in section \ref{sec:null}.
\subsection{Parameter and hypothesis priors ($\Hs$)}
\label{sec:priorsHs}
There are two types of prior probabilities that we need to compute in our signal hypothesis: $P(\teta|\Hs^{n})$ and $P(\Hs^{n})$. In order to find $P(\teta|\Hs^{n})$ we again use Bayes' rule
\begin{equation}
P(\teta|\Hs^{n})=\frac{P(\Hs^n|\teta)P(\teta)}{P(\Hs^n)}
\label{eq:4}
\end{equation}
\(P(\Hs^n)\) in the denominator and in Eq \ref{eq:1} cancel. So actually we need to have \(P(\teta)\) and \(P(\Hs^n|\teta)\).
We first discuss the prior probability distribution of the parameters, $P(\teta)$. Here we review the role of each source parameter.

\begin{enumerate}
\item {\bf Time ($t_{\rm s}$):} We assume that a signal is equally likely to occur at any time during an observation period. We further assume that no other parameter depends on the time of observation, therefore we can treat this probability independently. Taking the livetime duration of the joint observation period to be $T_{\rm obs}$, the resulting prior probability distribution is
\begin{equation}
P(t_{\rm s}) = \frac{1}{T_{\rm obs}}.
\label{eq:time}
\end{equation}
\item {\bf Source distance ($r$):} We assume a uniform distribution of sources in the universe \textcolor{black}{such that \(P(r)=3r^2/r_{max}^3\). \(r_{max}\) is the maximum value of \(r\) and its divergence is not important as it gets cancelled in the analysis.} We further assume that a GW signal can be detected if its root-sum-squared GW strain $h_{\rm rss}$ is above a detection threshold $h_{\rm rss,0}$ \cite{2010CQGra..27q3001A}. The probability density that an observed GW+neutrino event occurred at distance $r$ is dependent on $r$ since the volume in space in the distance range $[r,r+dr]$ is $\propto r^2 dr$; but the probability of detecting \(n\) neutrinos from the source falls according to Poisson probabilities for \(n\) detections whose means are proportional to $r^{-2}$. This dependency is valid up to the GW distance range $r_0 f_{\rm A}(\dir,t_{\rm gw})$ beyond which sources are not detected. Here, $r_0$ is the GW detection range for optimal source direction, and $f_{\rm A}(\dir,t_{\rm gw})$ is the antenna pattern of the GW detector network. \textcolor{black}{The latter is the square root of the quadrature sum of the antenna responses of each detector for the two polarizations of GWs as
\begin{equation}
f_A(\dir,t_{\rm gw})=\sqrt{\sum_k {F_{k,+}(\psi,\dir,t_{\rm gw})}^2+{F_{k,\times}(\psi,\dir,t_{\rm gw})}^2}
\end{equation} where sum over \(k\) is for summing over different detectors, \(F\) functions are the antenna responses for each polarization and \(\psi\) angle is the GW emission inclination which vanishes after the quadrature sum. }Its maximum value is 1 which corresponds to the optimal source position. The range $r_0$ satisfies $r_0(E_{\rm gw})\propto E_{\rm gw}^{1/2}$ and is defined as the distance at which an event which emits \(E_{\rm gw}\) amount of energy creates the least acceptable SNR \(\rho_{\rm min}\) or the strain \(h_{\rm rss,0}\). The relationship between \(E_{\rm gw},\rho_{\rm gw}\) and \(r_0\) is explained in Eq. \ref{eq:delta} with \(\rho_{\rm gw}=\rho_{\rm min}\)  and \(r=r_0\) such that \(r_0=\kappa_0 E_{\rm gw}^{1/2}/\rho_{\rm min}\). 

\item {\bf Energy ($E_{\rm gw}$ and $E_{\rm \nu}$):} We need to specify our dependency to energies. A naive choice can be independent log-uniform distributions over the energy ranges \([E_{\rm gw}^{-}, E_{\rm gw}^{+}], [E_{\rm \nu}^{-}, E_{\rm \nu}^{+}]\) with probability density
\begin{equation}
\begin{split}
P(E_{\rm gw},E_{\rm \nu})=P(E_{\rm gw})P(E_{\nu})\\=(E_{\rm gw}E_{\rm \nu}\log(\frac{E_{\rm gw}^{+}}{E_{\rm gw}^{-}})\log(\frac{E_{\rm \nu}^{+}}{E_{\rm \nu}^{-}}))^{-1}
\end{split}
\end{equation}
Throughout the paper instead of the expression for the specific log-uniform model, \(P(E_{\rm gw},E_{\nu})\) will be used to express the universality of the method.
\item {\bf Sky position (\(\dir\)):} We assume uniform prior distribution in the sky as \(P(\dir)=1/4\pi\)
\end{enumerate}
Overall, we find

\begin{equation}
P(\teta) =  \frac{P(E_{\rm gw},E_{\rm \nu})r^2}{4\pi T_{\rm obs}N_1}
\label{eq:parameterprior}
\end{equation}

with suitable normalization constant $N_1=r^3_{max}/3$ for maximum assumed possible distance \(r_{max}\). Its divergence is not important and all divergences in the analysis cancel.

Next we consider the term \(P(\Hs^n|\teta)\) which depends on the expected detection count of multi-messenger events.  A useful quantity for it is the expected number of detected neutrinos for given emission energy, sky location and distance per event
\begin{equation}
\langle n_{\nu}(E_{\rm \nu},r,\dir)\rangle = n_{\nu,51,100}(\dir) \left(\frac{E_{\nu}}{10^{51}\,\mbox{erg}}\right)\left(\frac{r}{100\,\mbox{Mpc}}\right)^{-2}
\end{equation}
Here, $n_{\nu,51,100}(\dir)$ is a detector specific parameter. Sky averaged $n_{\nu,51,100}\approx1.1$ for IceCube \cite{2018arXiv181011467B} and  $n_{\nu,51,100}(\dir)$ only depends on the declination but not to right ascension due to axial symmetry of the detector whose symmetry axis coincides with Earth's rotation axis due to its location at the South Pole. 

Probability of detection of multi-messenger events given the source parameters will be 
\begin{multline}
P_{det}^n(\teta)=\mbox{Poiss}(n,\langle n_{\nu}(E_{\nu},r,\dir)\rangle)\\ \begin{cases} 1 &  r\leq r_0(E_{\rm gw})\Bar{f}_A(\dir,t_s) \\ 0 & \text{otherwise}
\end{cases}
\end{multline}
with the time averaged antenna pattern \(\Bar{f}_A(\dir,t_s)\) between \([t_s+t_{\rm gw}^-,t_s+t_{\rm gw}^+]\). Parameters \(t_{\rm gw}^-\) and \(t_{\rm gw}^+\) will be explained in section \ref{sec:GWHs}. Poiss$(k,\lambda)$ is the Poisson probability density function with $\lambda$ mean and $k$ observed events. Then the expected count is 
\begin{equation}
C^n_{det}(\teta)=\dot{n}_{\rm gw+\nu}T_{obs}P^n_{det}(\teta)
\end{equation}
where $\dot{n}_{\rm gw+\nu}$ is the total multi-messenger event rate in the whole universe, which is bounded by the distance \(r_{max}\). Its divergence cancels the divergence of \(N_1\). Then the prior probability will be
\begin{equation}
P(\Hs^{n}|\teta) = \frac{1}{N_{2}} C^n_{det}(\teta)
\label{eq:Hs}
\end{equation}
with suitable normalization constant $N_{2}$. This constant, which is the sum of the expected event counts of all hypotheses, will be the present for our other hypotheses too and will be cancelled out.

\subsection{Gravitational waves ($\Hs$)}
\label{sec:GWHs}

We now consider the probability $P(\xgw|\teta, \Hs^{n})$.
We have:
\begin{multline}
P(\xgw|\teta, \Hs^{n}) = P(t_{\rm gw}|\teta,\Hs^{n})P(\rho_{\rm gw}|t_{\rm gw},\teta,\Hs^{n})\\ \times P(\mathcal{P}_{\rm gw}|t_{\rm gw},\rho_{\rm gw},\teta,\Hs^{n}) P(\mathcal{D}_{\rm gw}|\mathcal{P}_{\rm gw},t_{\rm gw},\rho_{\rm gw},\teta,\Hs^{n})
\label{eq:234243432}
\end{multline}

The term $P(t_{\rm gw}|\teta,\Hs^{n})$ should only depend on the difference $t_{\rm gw}-t_{\rm s}$. We adopt the assumption that the probability $P(t_{\rm gw}|t_{\rm s},\Hs^{n})$ is uniform within a time window $t_{\rm gw}-t_{\rm s} \in [t_{\rm gw}^{-},t_{\rm gw}^{+}]$ for suitable parameters $t_{\rm gw}^{-}$ and
$t_{\rm gw}^{+}$ and is zero elsewhere:
\begin{equation}
 P(t_{\rm gw}|t_{\rm s},\Hs^{n})=
 \begin{cases} (t_{\rm gw}^{+}-t_{\rm gw}^{-})^{-1} & \text{if } t_{\rm gw}-t_{\rm s}\in [t_{\rm gw}^{-},t_{\rm gw}^{+}]  \\
                      0                                    & \mbox{otherwise}
 \end{cases}
\end{equation}
For example, previous GW+neutrino searches used parameters $t_{\rm gw}^{+}=-t_{\rm gw}^{-}=250$\,s \cite{2011APh....35....1B,2012PhRvD..85j3004B,2014PhRvD..90j2002A,2016PhRvD..93l2010A,2017PhRvD..96b2005A,2017ApJ...850L..35A}. We assume that the other source parameters are independent of $t_{\rm gw}$.

To understand the second term on the right hand side of Eq. \ref{eq:234243432}, we make use of the fact that $\rho_{\rm gw}$ on average is proportional to the GW signal's amplitude at Earth, characterized by the root-sum-squared GW strain $h_{\rm rss}$. Assuming here for simplicity that all gravitational waveforms are similar, the GW strain is fully determined by $r$, $E_{\rm gw}$, $\dir$ and $t_{\rm gw}$. The time dependence comes in due to Earth's rotation if we measure sky location in equatorial coordinates. Assuming that $\rho_{\rm gw}$ precisely describes $h_{\rm rss}$, this term represents a constraint on the source parameters, which need to be such that they produce at Earth the $h_{\rm rss}$ value that corresponds to the measured $\rho_{\rm gw}$ value. This means that only the combination $E_{\rm gw}^{1/2}r^{-1}f_{\rm A}(\dir,t_{\rm gw})$ is constrained, where $f_{\rm A}(\dir,t_{\rm gw})$ is the direction-dependent antenna pattern of the GW detector network. This combination is proportional to the measured GW strain amplitude. 

We therefore write the probability as a constraint
\begin{equation}
\label{eq:delta}
P(\rho_{\rm gw}|t_{\rm gw},\teta,\Hs^{n}) = \delta\left[\rho_{\rm gw} - \kappa_0 E_{\rm gw}^{1/2}r^{-1}f_{\rm A}(\dir,t_{\rm gw})\right]
\end{equation}
Where $\delta$ is the Dirac-delta and $\kappa_0$ is an appropriate constant that depends on the noise spectrum in the GW detector at the time of detection and on the GW search algorithm. The delta distribution is an approximation for the accurate measurement of the SNR although in practice there is always an uncertainty in the measured SNR. We do not consider that uncertainty in our analysis.

Next, we look at the term $P(\mathcal{P}_{\rm gw}|t_{\rm gw},\rho_{\rm gw},\teta,\Hs^{n})$. Using Bayes' rule, we write
\begin{multline}
P(\mathcal{P}_{\rm gw}|t_{\rm gw},\rho_{\rm gw},\teta,\Hs^{n}) \\ = \frac{P(\teta|\mathcal{P}_{\rm gw},t_{\rm gw},\rho_{\rm gw},\Hs^{n})P(\mathcal{P}_{\rm gw}|t_{\rm gw},\rho_{\rm gw},\Hs^{n})}{P(\teta|t_{\rm gw},\rho_{\rm gw},\Hs^{n})}
\label{eq:45645465}
\end{multline}
Regarding $P(\mathcal{P}_{\rm gw}|t_{\rm gw},\rho_{\rm gw},\Hs^{n})$, we assume that the distribution of $\mathcal{P}_{\rm gw}$ is independent of the underlying hypothesis, i.e. $P(\mathcal{P}_{\rm gw}|t_{\rm gw},\rho_{\rm gw},\Hs^{n}) = P(\mathcal{P}_{\rm gw}|t_{\rm gw},\rho_{\rm gw})$. This term appears for the alternative hypothesis as well, it cancels out and therefore we can ignore it here. For the remaining terms assuming that our reconstructed GW skymap is accurate, we have
\begin{equation}
\frac{P(\teta|\mathcal{P}_{\rm gw},t_{\rm gw},\rho_{\rm gw},\Hs^{n})}{P(\teta|t_{\rm gw},\rho_{\rm gw},\Hs^{n})} = \mathcal{P}_{\rm gw}(\dir)\frac{N_\dir}{f_A(\dir,t_{\rm gw})}
\end{equation}
The numerator of the left side gives the skymap since the skymap determines the probability of the signal coming from a given sky location $\dir$. The denominator of the left side is proportional to the antenna pattern at the time of detection, normalized by the factor \(N_\dir\)

Now we look at the term $P(\mathcal{D}_{\rm gw}|\mathcal{P}_{\rm gw},t_{\rm gw},\rho_{\rm gw},\teta,\Hs^{n})$. The handling of this term is identical to $P(\mathcal{P}_{\rm gw}|t_{\rm gw},\rho_{\rm gw},\teta,\Hs^{n})$. Again we assume the independence of \(\mathcal{D}_{\rm gw}\) and the hypothesis, and write
\begin{multline}
P(\mathcal{D}_{\rm gw}|\mathcal{P}_{\rm gw},t_{\rm gw},\rho_{\rm gw},\teta,\Hs^{n}) \\ =   \frac{P(\teta|\mathcal{P}_{\rm gw},t_{\rm gw},\rho_{\rm gw},\mathcal{D}_{\rm gw},\Hs^{n})P(\mathcal{D}_{\rm gw}|\mathcal{P}_{\rm gw},t_{\rm gw},\rho_{\rm gw},\Hs^{n})}{P(\teta|\mathcal{P}_{\rm gw},t_{\rm gw},\rho_{\rm gw},\Hs^{n})}
\label{eq:dist1}
\end{multline}
We again ignore the second term in the denominator as it cancels with the corresponding terms in other hypotheses. Assuming that our reconstructed GW distance distribution is accurate, we have
\begin{equation}
\frac{P(\teta|\mathcal{D}_{\rm gw},\mathcal{P}_{\rm gw},t_{\rm gw},\rho_{\rm gw},\Hs^{n})}{P(\teta|\mathcal{P}_{\rm gw}, t_{\rm gw},\rho_{\rm gw},\Hs^{n})}=\mathcal{D}_{\rm gw}(r)\frac{N_r(\rho_{\rm gw})}{r^2\times r^{-1}}
\end{equation}
The first term in the right side comes from the numerator of the left side since the distance distribution determines the probability of the signal coming from a source distance $r$. The second term in the right side represents the denominator of the left side and is obtained by multiplying the prior \(r^2\) distribution and the \(r^{-1}\) distribution of the likelihood of \(r\) for known \(\rho_{\rm gw}\) which is proportional to the distribution of \(E_{\rm gw}^{1/2}\) which is \(\propto E_{\rm gw}^{-1/2}\) for a log uniformly assumed \(P(E_{\rm gw})\). Here \(N_r(\rho_{\rm gw})\) is the normalization for the \(r\) distribution between \([\sqrt{E^-_{\rm gw}},\sqrt{E^+_{\rm gw}}]\rho_{\rm gw}/\kappa_0\).

Putting everything together, we have for the GW term without the cancelling terms
\begin{widetext}
\begin{equation}
P(\xgw|\teta, \Hs^{n}) = N_\dir N_r(\rho_{\rm gw}) \delta\left[\rho_{\rm gw} - \kappa_0 E_{\rm gw}^{1/2}r^{-1}f_{\rm A}(\dir,t_{\rm gw})\right] \frac{\mathcal{P}_{\rm gw}(\dir)}{f_A(\dir,t_{\rm gw})}\frac{\mathcal{D}_{\rm gw}(r)}{r}
\begin{cases} (t_{\rm gw}^{+}-t_{\rm gw}^{-})^{-1} & \text{if } t_{\rm gw}-t_{\rm s}\in [t_{\rm gw}^{-},t_{\rm gw}^{+}]  \\
                      0                                    & \mbox{otherwise}
\end{cases}
\label{eq:gwsignal}
\end{equation}
\end{widetext}

\subsection{High-energy neutrinos ($\Hs$)}

We now turn our attention to the high-energy neutrino term $P(\xnui,\xnuj ...|\teta, \Hs^{n})$ as in Eq. \ref{eq:3}. We assume no dependency of the observables of different neutrinos on other neutrinos' observables, except the dependency through \(\teta\). Therefore we can separate each neutrino terms as 
\begin{equation}
P(\xnui,\xnuj ...|\teta, \Hs^{n})=P(\xnui|\teta,\Hs^{n})P(\xnuj|\teta,\Hs^{n})...
\end{equation}
We treat the temporal term similarly to the GW case. We assume that the time difference $t_{\rm \nu i}-t_{\rm s}$ is the only relevant temporal value. We further use a uniform probability density within the time interval $[t_{\nu}^{-}$,$t_{\nu}^{+}]$, and 0 outside:
\begin{equation}
 P(t_{\nu i}|t_{\rm s},\Hs^{n})=
 \begin{cases} (t_{\nu}^{+}-t_{\nu}^{-})^{-1} & \text{if } t_{\rm \nu i}-t_{\rm s}\in [t_{\nu}^{-},t_{\nu}^{+}]  \\
                      0                                    & \mbox{otherwise}
 \end{cases}
.
\end{equation}
Previous GW+neutrino searches used parameters $t_{\nu}^{+}=-t_{\nu}^{-}=250$\,s \cite{2011APh....35....1B,2012PhRvD..85j3004B,2014PhRvD..90j2002A,2016PhRvD..93l2010A,2017PhRvD..96b2005A,2017ApJ...850L..35A}.

The remaining neutrino observables, $\dir_{\rm \nu i}$, $\sigma_{\rm \nu i}$ and $\epsilon_{\rm \nu i}$, are not independent. The sensitivity of neutrino detectors varies with both energy and sky location, and localization accuracy depends on source direction and energy. 

Let us take the remaining neutrino term \(P(\dir_{\rm \nu i}, \sigma_{\rm \nu i}, \epsilon_{\rm \nu i} | r, \dir, E_{\nu}, \Hs^{n})\). We assume that the signal distribution of $\epsilon_{\nu}$ follows a power law, therefore the neutrino spectrum is independent of the source distance. Such a power-law distribution is typical in neutrino emission models \cite{1997PhRvL..78.2292W}. Consequently, parameters $r$ and $E_{\nu}$ do not affect the probability here. We further assume that the directional uncertainty variable $\sigma_{\nu}$ and the reconstructed sky position of neutrino \(\dir_{\nu}\) don't depend on \(r\) and \(E_{\nu}\).

We use the chain rule to write
\begin{widetext}
\begin{equation}
P(\dir_{\rm \nu i}, \epsilon_{\rm \nu i }, \sigma_{\rm \nu i}| \dir, \Hs^{n}) = P(\dir_{\rm \nu i} | \sigma_{\rm \nu i}, \epsilon_{\rm \nu i}, \dir, \Hs^{n})P(\sigma_{\rm \nu i}|\epsilon_{\rm \nu i},\dir, \Hs^{n})P(\epsilon_{\rm \nu i}|\dir, \Hs^{n})
\label{eq:76567}
\end{equation}
\end{widetext}
Given the source direction as a parameter, the probability of reconstructing $\epsilon_\nu$ for a detected neutrino depends on the energy- and direction-dependent effective area $A_{\rm eff}(\epsilon_{\nu},\dir)$ of the neutrino detector, as well as the source power spectral density. Here we ignore the difference between true and reconstructed energy when calculating the effective area as this should not significantly change its value. We take the neutrino spectral density to be $dN_{\nu}/d\epsilon_{\nu}\propto \epsilon_{\nu}^{-2}$, which is the standard spectrum expected from Fermi processes \cite{1997PhRvL..78.2292W}. With these dependencies, we write
\begin{equation}
P(\epsilon_{\rm \nu i} | \dir, \Hs^{n}) = \frac{1}{N_{\epsilon}}A_{\rm eff}(\epsilon_{\rm \nu i},\dir)\epsilon_{\rm \nu i}^{-2},
\end{equation}
where 
\begin{equation}
N_{\epsilon}=\int d\dir \int_{\epsilon_{\rm min}}^{\epsilon_{\rm max}} \epsilon_{\nu}^{-2}A_{\rm eff}(\epsilon_{\nu},\dir)d\epsilon_{\nu}
\end{equation}
where the \(d \dir\) integral is over all sky and \(\epsilon_{\rm min}\) and \(\epsilon_{\rm max}\) are the minimum and maximum reconstructable energies.

We assume \(P(\sigma_{\rm \nu i}|\epsilon_{\rm \nu i})\) does not depend on the hypothesis of consideration or its parameters; therefore will be cancelled with the same term in the other hypotheses when the ratio of probabilities is taken at the very end. Hence we do not consider the actual value of  \(P(\sigma_{\rm \nu i}|\epsilon_{\rm \nu i},\dir,\Hs^{n})\).

For the first term on the right-hand side in Eq. \ref{eq:76567}, we adopt the Normal distribution \cite{2008APh....29..299B}
\begin{equation}
P(\dir_{\rm \nu i} | \sigma_{\rm \nu i}, \epsilon_{\rm \nu i}, \dir, \Hs^{n}) = \frac{1}{2\pi\sigma_{\rm \nu i}^{2}}\mbox{e}^{-\frac{|\dir_{\rm \nu i} - \dir|^2}{2\sigma_{\rm \nu i}^{2}}}
\end{equation}
by assuming no further dependence for \(\dir_{\rm \nu i}\) on \(\epsilon_{\rm \nu i}\) except the one through \(\dir\). Putting everything together, we have for the neutrino term
\begin{widetext}
\begin{equation}
P(\xnui|\teta, \Hs^{n}) = \frac{1}{N_{\epsilon}}A_{\rm eff}(\epsilon_{\rm \nu i},\dir)\epsilon_{\rm \nu i}^{-2} \frac{1}{2\pi\sigma_{\rm \nu i}^{2}}\mbox{e}^{-\frac{|\dir_{\rm \nu i} - \dir|^2}{2\sigma_{\rm \nu i}^{2}}}
\begin{cases} (t_{\nu}^{+}-t_{\nu}^{-})^{-1} & \text{if } t_{\rm \nu i}-t_{\rm s}\in [t_{\nu}^{-},t_{\nu}^{+}]  \\
                      0                                    & \mbox{otherwise}
\end{cases}
\label{eq:neutrinosignal}
\end{equation}
\end{widetext}

\subsection{Combination of probabilities ($\Hs$)}

We can combine the above results to obtain the probability of the joint event being a signal by taking Eqs. \ref{eq:gwsignal} and \ref{eq:neutrinosignal} and substituting them into Eq. \ref{eq:3}. We then substitute Eq. \ref{eq:3} into Eq. \ref{eq:2}. 

To solve Eq. \ref{eq:2}, we further substitute $P(\teta|\Hs^{n})$ from Eq. \ref{eq:4} for which we use Eqs. \ref{eq:parameterprior} and \ref{eq:Hs}. Then, we can substitute Eq. \ref{eq:2} into Eq. \ref{eq:1.6}. Finally we substitute Eq.  \ref{eq:1.6} into Eq. \ref{eq:1.5} with Eq. \ref{eq:1.55} which is the required term for Eq. \ref{eq:1}, where we obtain $P(\Hs^{n}|\xgw, \Xnu)$, except the factor $P(\xgw,\Xnu)$ which will cancel out in comparison to the alternative hypothesis. Computation of \(P(\Xnu^{i,j,...}|\Ho)\) will be explained in section \ref{sec:null}.

\section{Null hypothesis}
\label{sec:null}

We now move to our null hypothesis $\Ho$. Given the observational data, the probability of the null hypothesis being true can be written as $P(\Ho|\xgw, \Xnu)$. We apply Bayes' rule to express this probability as
\begin{equation}
P(\Ho|\xgw, \Xnu) = \frac{P(\xgw, \Xnu|\Ho)P(\Ho)}{P(\xgw,\Xnu)},
\label{eq:1b}
\end{equation}
The denominator here will cancel out with the same denominator in the signal hypothesis, therefore we do not need to further consider it.
Since the background events for GW and neutrino observations are independent, we can write 
\begin{equation}
P(\xgw, \Xnu|\Ho) = P(\xgw|\Ho)P(\Xnu|\Ho)
\label{eq:3b}
\end{equation}
We will now specify the independent elements of Eqs. \ref{eq:1b} and \ref{eq:3b} in the context of our background model. We can perform the calculations in this section without a need for additional parameters to marginalize over due to the fact that every measured parameter is assumed to be independent of each other for background.

\subsection{Hypothesis prior ($\Ho$)}

There is one prior probability that we need to compute in our null hypothesis: $P(\Ho)$. This probability is again proportional to the expected detection count of background events. Given the observation period \(T_{obs}\) and the background GW and rate \(R_{\rm gw,bg}\), we have \(T_{obs}R_{\rm gw,bg}\) background GW events. Hence
\begin{equation}
P(\Ho) = \frac{1}{N_{2}}R_{\rm gw,bg}T_{obs}
\label{eq:pH0}
\end{equation}
The normalization factor $N_{2}$ will cancel out with the same factor in the signal hypothesis, see Eq. \ref{eq:Hs}. 

\subsection{Gravitational waves ($\Ho$)}

We now consider the GW component $P(\xgw|\Ho)$. We assume \(t_{\rm gw}\) and \(\rho_{\rm gw}\) are independent. We can then define the probabilities of measuring each parameter independently:
\begin{multline}
P(\xgw| \Ho) =
P(t_{\rm gw}| \Ho) P(\rho_{\rm gw}|\Ho)\\ \times P(\mathcal{P}_{\rm gw}|t_{\rm gw},\rho_{\rm gw},\Ho)P(\mathcal{D}_{\rm gw}|t_{\rm gw},\rho_{\rm gw},\mathcal{P}_{\rm gw},\Ho)
\end{multline}

Starting with the first term on the right-hand side, we expect the probability distribution of detection time for a background event to be independent of time, therefore we adopt a uniform distribution within the observation time. We therefore have
\begin{equation}
 P(t_{\rm gw}|\Ho)=\frac{1}{T_{obs}}
 \label{eq:H0gwtime}
\end{equation}

The distribution of $\rho_{\rm gw}$ depends on the detector properties as well as the properties of the reconstruction algorithm. We therefore estimate this distribution empirically. \textcolor{black}{We use the $\rho_{\rm gw}$ background distribution obtained from the GW search pipelines which produce it by time shifting data between multiple GW observatories and carrying out the full analysis algorithm over this time shifted data. We do not extrapolate beyond the maximum $\rho_{\rm gw}$ from the background distribution and conservatively choose it as the highest value for the background triggers.} We denote the empirically established distribution of $\rho_{\rm gw}$ with $P_{\rm emp}(\rho_{\rm gw}|\Ho)$. The \(\rho_{\rm gw}\) obtained by time shifts will be acquired from the GW data analysis pipelines.

Considering the terms $P(\mathcal{P}_{\rm gw}|t_{\rm gw},\rho_{\rm gw},\Ho)$ and $P(\mathcal{D}_{\rm gw}|t_{\rm gw},\rho_{\rm gw},\mathcal{P}_{\rm gw},\Ho)$, we do not have any prior information on $P(\mathcal{P}_{\rm gw}|\Ho)$ and $P(\mathcal{D}_{\rm gw}|\Ho)$, therefore we assume that it is independent of $\mathcal{P}_{\rm gw}$ and $\mathcal{D}_{\rm gw}$. Since there are similar terms in our signal hypothesis, these cancel out.  We therefore ignore these terms in the following.

Putting everything together, we have for the background GW term
\begin{equation}
P(\xgw| \Ho) = P_{\rm emp}(\rho_{\rm gw}|\Ho)\frac{1}{T_{obs}}
\label{eq:gwsignalH0}
\end{equation}

\subsection{High-energy neutrinos ($\Ho$)}

Next, we examine $P(\Xnu|\Ho)$ in Eq. \ref{eq:3b} and also \(P(\Xnu^{i,j,...}|\Ho)\) in Eq. \ref{eq:1.6}. Given the background neutrino rate \(R_{\rm \nu, bg}\), the probability of having N background neutrinos in the observation period is \({\rm Poiss}(N,R_{\rm \nu, bg}T_{obs})\), which allows us to write
\begin{equation}
 P(\Xnu|\Ho)=P(\Xnu|\Ho,\#_{bg}=N)P(\#_{bg}=N|\Ho)
 \label{eq:sumbg}
\end{equation}
with \(P(\#_{bg}=N|\Ho) ={\rm Poiss}(N,R_{\rm \nu, bg}T_{obs})\) where \(\#_{bg}\) is the number of background neutrinos. For short notation we will use \(\Ho^N\) instead of both \(\Ho\) and \(\#_{bg}=N\). Now we can decompose the first term into single neutrino terms as we did previously in section \ref{sec:sig}. Then we first separate the temporal term which we assume to be independent of the other parameters. We assume that the time of arrival of a background neutrino signal is time-independent, and can be anytime during the observation period. We therefore have for each neutrino
\begin{equation}
 P(t_{\nu i}|\Ho^N)=\frac{1}{{T_{obs}}}
 \label{eq:H0nutime}
\end{equation}

The remaining measured parameters will not be independent of each other. In particular, the reconstructed neutrino direction and energy are interconnected. As we explained before the term for directional uncertainty parameter $ \sigma_{\nu i}$ will be cancelled when we decompose the remaining terms as
\begin{multline}
P(\dir_{\rm \nu i}, \epsilon_{\rm \nu i }, \sigma_{\rm \nu i}| \Ho^N) = \\ P(\dir_{\rm \nu i} | \sigma_{\rm \nu i}, \epsilon_{\rm \nu i}, \Ho^N)P(\sigma_{\rm \nu i}|\epsilon_{\rm \nu i},\Ho^N)P(\epsilon_{\rm \nu i}|\Ho^N)
\end{multline}
In addition we assume \(P(\dir_{\rm \nu i} | \sigma_{\rm \nu i}, \epsilon_{\rm \nu i}, \Ho^N)\) term doesn't have any \(\sigma_{\nu i}\) dependence. We therefore effectively need to examine the probability $P(\dir_{\rm \nu i} | \epsilon_{\rm \nu i}, \Ho^N)P(\epsilon_{\rm \nu i}| \Ho^N)= P(\dir_{\nu},\epsilon_{\nu}|\Ho^N)$. Given a sufficient number of observed background events, this probability can be estimated empirically using observed data. Let $\{\dir_{\rm \nu,j},\epsilon_{\rm \nu,j}\}, j\in N_{\rm \nu,obs}$ be the reconstructed parameters observed set of $N_{\rm \nu,obs}$ neutrino candidates. For the direction we only care about the declination angle in the equatorial coordinate system \(\delta_{\nu i}(\dir_{\nu i})\) primarily because of the axial symmetry for IceCube, which is described in section \ref{sec:priorsHs} when commenting on \(n_{\rm \nu,51,100}(\dir)\). For non-coaxial detectors with Earth's rotation axis, full \(\dir\) should be considered. We then have the empirical estimate with the kernel density estimation
\begin{widetext}
\begin{equation}
P_{\rm emp}(\dir_{\nu i},\epsilon_{\nu i}|\Ho^N) =\frac{\sum_{j\in N_{\rm \nu,obs}} \left[ |\delta_{\nu i} - \delta_{\rm \nu,j}| < \Delta_\delta \,\,\,\,\&\,\,\,\, |\epsilon_{\rm \nu i} - \epsilon_{\rm \nu,j}| < \Delta_\epsilon(\epsilon_{\rm \nu i})  \right]}{4\pi N_{\rm \nu,obs} |cos(\delta_{\nu i}+\Delta_\delta)-cos(\delta_{\nu i}-\Delta_\delta)|\Delta_\epsilon(\epsilon_{\rm \nu i})}
\end{equation}
\end{widetext}
where we use here the bracket notation such that $[P]$ is $1$ if $P$ is true and $0$ if $P$ is false which corresponds to the top hat kernel. We further introduced constants $\Delta_\delta$ and $\Delta_\epsilon(\epsilon_{\rm \nu})$, which should be selected such that the uncertainty on the probability estimate is minimal.

Putting everything together, we have for the background neutrino term
\begin{equation}
P(\xnui| \Ho^N) = P_{\rm emp}(\dir_{\nu i},\epsilon_{\nu i}|\Ho^N)\frac{1}{T_{obs}}
\label{eq:neutrinoH0full}
\end{equation}

\subsection{Combination of probabilities ($\Ho$)}

We can combine the above results to obtain the probability of the joint event being from the background by taking Eq. \ref{eq:neutrinoH0full} for each neutrino and substituting them in Eq. \ref{eq:sumbg}. Then Eq. \ref{eq:gwsignalH0} and Eq. \ref{eq:sumbg} can be substitued into Eq. \ref{eq:3b}. Finally Eq. \ref{eq:3b} along with Eq. \ref{eq:pH0} can be substituted into Eq. \ref{eq:1b}. Eq. \ref{eq:1b} will miss a normalization factor both from Eq. \ref{eq:pH0} and from the denominator on the right side, both of which cancel out upon calculating the Bayes factor. For the background terms in other hypotheses such as \(P(\Xnu^{i,j,...}|\Ho)\) in Eq. \ref{eq:1.6}, Eq. \ref{eq:sumbg} can be used similarly for \(N-n\) number of background neutrinos instead of \(N\).

\section{Chance coincidence hypothesis}
\label{sec:coincidence}
We finally calculate the probability for the chance coincidence hypothesis $\Hc$. Given the observational data the probability of the chance coincidence hypothesis being true can be written as $P(\Hc|\xgw, \Xnu)$. $\Hc$ can be separated into two parts, one of which considers a background neutrino event and a foreground gravitational wave event denoted by $H_{\rm c, gw}$; and the other one considers a background gravitational wave event and a foreground neutrino event denoted by $\Hcnu$. Since these two cases are mutually exclusive and complementary to each other for the chance coincidence hypothesis we can write \(P(\Hc|\xgw, \Xnu) = P(H_{\rm c, gw}|\xgw, \Xnu) + P(\Hcnu|\xgw, \Xnu)\). \(H_{\rm c, gw}\) is a special case for the signal hypothesis \(\Hs^n\) with \(n=0\), so all the explanations in section \ref{sec:sig} apply for it. Due to absence of related events, we have a simpler case. We apply Bayes' rule again.
\begin{equation}
P(\Hcgw|\xgw, \Xnu)= \frac{P(\xgw, \Xnu|\Hcgw)P(\Hcgw)}{P(\xgw, \Xnu)}
\end{equation}
\(P(\xgw, \Xnu)\) is omitted as it is through the paper. Then we separate the first term in the numerator due to independent detections as 
\begin{equation}
P(\xgw, \Xnu|\Hcgw)=P(\xgw|\Hcgw)P(\Xnu|\Hcgw)
\end{equation}
Then we obtain the GW part by marginalizing over parameters \(\teta\)
\begin{multline}
P(\xgw|\Hcgw)=\\ \int P(\xgw|\teta, \Hcgw)P(\teta|\Hcgw)d\teta
\end{multline}
with
\begin{equation}
    \teta=\{t_s, r,\dir,E_{\rm gw},E_{\nu}\}
\end{equation}
which are the same parameters defined in Section \ref{sec:sig}.
Now we move on analyzing \(P(\Hcnu|\xgw, \Xnu)\). We first decompose it in sub-hypotheses for different number of signal neutrinos, denoted as \(\Hcnu^n\) for \(n\) signal neutrinos.
\begin{equation}
P(\Hcnu|\xgw, \Xnu)=\sum_{n=1}^N P(\Hcnu^n|\xgw, \Xnu)
\end{equation}
We again apply Bayes' rule to each term
\begin{equation}
\label{eq:1c}
P(\Hcnu^n|\xgw, \Xnu)= \frac{P(\xgw, \Xnu|\Hcnu^n)P(\Hcnu^n)}{P(\xgw, \Xnu)}
\end{equation}
Here the denominator again cancels out with the same denominator in the signal and null hypotheses. We again separate the first term in numerator due to independent detections as 
\begin{equation}
P(\xgw, \Xnu|\Hcnu^n)=P(\xgw|\Hcnu^n)P(\Xnu|\Hcnu^n)
\end{equation}
We further decompose the neutrino term by identifying the signal neutrinos with the set \(s\) which has \(n\) elements as in Section \ref{sec:sig}
\begin{widetext}
\begin{equation}
P(\Xnu|\Hcnu^n)=\sum_{\{i,j,...\}}P(\Xnu|\Hcnu^n,s=\{i,j,...\})P(s=\{i,j,...\}|\Hcnu^n)
\end{equation}
\end{widetext}
with 
\begin{equation}
P(s=\{i,j,...\}|\Hcnu^n)={N \choose n}^{-1}
\end{equation}
We separate the signal and background neutrinos due to independence of different detections and drop the set \(s\) in notation as we did in Eq \ref{eq:1.6}
\begin{multline}
P(\Xnu|\Hcnu^n,s=\{i,j,...\})=\\ P(\xnui,\xnuj,...|\Hcnu^n)P(\Xnu^{i,j,...}|\Ho)
\end{multline}
The second term was explained in Section \ref{sec:null}. In order to obtain the first term we marginalize over parameters \(\teta\)
\begin{equation}
\label{eq:2c}
\begin{split}
P(\xnui,\xnuj...|\Hcnu^n)= \\
\int P(\xnui,\xnuj,...|\teta, \Hcnu^n)P(\teta|\Hcnu^n)d\teta
\end{split}
\end{equation}
We again split each neutrino as we did in Section \ref{sec:sig}
\begin{equation}
P(\xnui,\xnuj,...|\teta, \Hcnu^n)=P(\xnui|\teta, \Hcnu^n)P(\xnuj|\teta, \Hcnu^n)...
\end{equation}
\subsection{Parameter and hypothesis priors (${\boldsymbol {\Hcgw}}$)}
For \(P(\teta|\Hcgw)\) we use Bayes' rule.
\begin{equation}
P(\teta|\Hcgw)=\frac{P(\Hcgw|\teta)P(\teta)}{P(\Hcgw)}
\end{equation}
with \textcolor{black}{\(P(\teta)\) in Eq. \ref{eq:parameterprior}.}

We write the probability density for detecting a GW event but not a neutrino as 
\begin{multline}
P_{det}^{\rm c, gw}(\teta, \alpha)=(\alpha \mbox{Poiss}(0,\langle n_{\nu}(E_{\nu},r,\dir)\rangle)+(1-\alpha))\\ \begin{cases} 1 &  r\leq r_0(E_{\rm gw})\Bar{f}_A(\dir,t_s) \\ 0 & \text{otherwise}
\end{cases}
\end{multline}
with \(\alpha\) being the ratio of total multi-messenger event rate to total astrophysical GW event rate. Then
\begin{equation}
P(\Hcgw|\teta)=\frac{1}{N_2}T_{obs}\dot{n}_{\rm gw}P_{det}^{\rm c,gw}(\teta,\alpha)
\end{equation}
with \(\dot{n}_{\rm gw}\) being the total astrophysical GW event rate in the whole universe.
\subsection{Parameter and hypothesis priors (${\boldsymbol {\Hcnu}}$)}
For \(P(\teta|\Hcnu^n)\) we use Bayes' rule.
\begin{equation}
P(\teta|\Hcnu^n)=\frac{P(\Hcnu^n|\teta)P(\teta)}{P(\Hcnu^n)}
\end{equation}
with \textcolor{black}{\(P(\teta)\) in Eq. \ref{eq:parameterprior}.}

We write the probability density for detecting \(n\) neutrinos but not a GW event as 
\begin{widetext}
\begin{equation}
P_{det}^{\rm c, \nu,n}(\teta, \beta)= \begin{cases} (1-\beta) &  r\leq r_0(E_{\rm gw})\bar{f}_A(\dir,t_s) \\ \beta \mbox{Poiss}(n,\langle n_{\nu}(E_{\nu},r,\dir)\rangle)+(1-\beta) & \text{otherwise}
\end{cases}
\end{equation}
\end{widetext}
with \(\beta\) being the ratio of total multi-messenger event rate to total astrophysical neutrino event rate. Then
\textcolor{black}{
\begin{equation}
P(\Hcnu^n|\teta)=\frac{1}{N_2}T_{obs}^2R_{\rm bg,gw}\dot{n}_{\rm \nu}P_{det}^{\rm c,\nu, n}(\teta,\beta)
\end{equation}}
with \(\dot{n}_{\nu}\) being the total astrophysical neutrino event rate in the universe.
\textcolor{black}{
\subsection{Remaining terms}
\begin{itemize}
\item The term $P(\xgw|\teta, \Hc^{\rm gw})$ is equal to the same term for our signal hypothesis, i.e. $P(\xgw|\teta, \Hs^n)$ (see Eq. \ref{eq:gwsignal}), since in both cases there is a detected astrophysical gravitational wave signal.
\item The term $P(\xnui|\teta, \Hcgw)$ is equal to the same term for our null hypothesis, i.e. $P(\xnui|\Ho)$ (see Eq. \ref{eq:neutrinoH0full}), since in both cases there is a background neutrino event, and neither term depends on the GW signal.
\item The term $P(\xgw|\Hcnu^n)$ is equal to the same term for our background hypothesis, i.e. $P(\xgw|\Ho)$ (see Eq. \ref{eq:gwsignalH0}), since in both cases there is a GW false detection from the background, and neither term depends on the neutrino signal.
\item The term $P(\xnui|\teta, \Hcnu^n)$ is equal to the same term for our signal hypothesis, i.e. $P(\xnui|\teta, \Hs^n)$ (see Eq. \ref{eq:neutrinosignal}), since in both cases there is a detected astrophysical neutrino, and neither term depends on the GW signal.
\end{itemize}
Combination of probabilities are done similarly to the signal and null hypothesis cases.}

\section{Odds ratio}
\label{sec:odds}
We test our signal hypothesis using odds ratios. We compare our signal hypothesis against both null and coincident hypotheses as in Eq. \ref{eq:or}.

It should be noted that this end result doesn't depend on \(T_{obs}\) which is a quantity fixed by humans' decisions and expected not to affect the significance of any astrophysical event. In addition there is no terms with explicit \(N\) dependency as expected since it could be arbitrarily large due to linear dependence on \(T_{obs}\). However, overall there is still a dependency due to the maximum possible signal neutrino count. In other words, \(N\) dependency of all terms got cancelled; but the number of terms depend on it.

This comparison will be applicable both for (i) GW and neutrino candidates that are not independently established detections, and for (ii) detections that are already confirmed through one channel. For the prior case, the first term in the denominator will be relevant, while in the latter case it will be the second term.

Although the odds ratio can be converted to a Bayesian probability for having a signal given the observations, it will be dependent on the parameter priors and the event rate densities which can be very uncertain. Therefore the odds ratio can be used as a test statistic. We empirically characterize the required threshold values based on background data and simulations for frequentist significances, similarly to \cite{Albert_2019}. For the searches with confirmed GW detections, the simulations consist of randomly paired simulated GWs and background neutrinos from previous detections. The number of background neutrinos in the time window around a GW is determined by Poisson distribution whose mean is the actual background neutrino rate times the duration of the time window. For searches with non-established GW detections, namely subthreshold searches, besides the GW and neutrino pairs for the previous case, there are pairs of time shifted background GW detections, which are acquired from GW data analysis pipelines, and background neutrinos, and pairs of background GW detections and signal neutrinos as well. All pairs are mixed in proportion to their estimated rates. These background comparisons allow us to determine a false alarm probability, namely p-value or significance, for the given events, which can be reported to initiate electromagnetic follow-up observations. 

\textcolor{black}{During the O3 public alert search for coincident GW and HENs the following parameter values have been used: \(t_{gw}^+=-t_{gw}^-=t_{\nu}^+=-t_{\nu}^-=250s,\ \dot{n}_{\rm gw+\nu}=\dot{n}_{\rm gw}=(4\pi r_{max}^3/3)1500\ Gpc^{-3}year^{-1},\ \alpha=1,\ R_{\rm \nu,bg}=6.4\times 10^{-3} Hz,\ E_{\nu}^+=10^{51}\ ergs,\ E_{\nu}^-=10^{46}\ ergs,\ \Delta_{\delta}=2.5^\circ,\ \Delta_{\epsilon}(\epsilon_{\nu})=0.3\times \epsilon_{\nu}\). Furthermore since it is a public alert search for verified GW detections. we set \(P(\Ho)=P(\Hcnu)=0\).}

\section{Conclusions}
\label{sec:conclusion}

We presented a search algorithm for common sources of GWs and high-energy neutrinos based on Bayesian hypothesis testing. This algorithm upgrades the method of Baret et al. \cite{2012PhRvD..85j3004B} that was used in most prior joint searches. The main advantages of the new method are that (i) it incorporates astrophysical priors about the source that help differentiate between signal and background, while being largely independent of the specific astrophysical model in consideration; (ii) it incorporates a more realistic model of the detector background, for example by taking into account the direction dependent background rate and energy distribution. These detector properties are straightforward to establish empirically, and the method presents a straightforward way to incorporate them as priors.

In the presentation of the method, we made simplifications that make the algorithm easier to implement and can make the computation faster. As an example, we assumed that all GW and neutrino sources emit the same energy. It will be useful to study how these simplifications affect the sensitivity of the search, and how much model dependence they introduce. This will be carried out in a future work.

\section*{Acknowledgments}
The authors are grateful for the useful feedback of Marek Szczepanczyk, Thomas Dent, Xilong Fan, and the IceCube Collaboration. The article has been approved for publication by the LIGO Scientific Collaboration under document number LIGO-P1800303. The authors thank the University of Florida and Columbia University in the City of New York for their generous support.
The Columbia Experimental Gravity group is grateful for the generous support of the National Science Foundation under grant PHY-1708028. \textcolor{black}{DV is grateful to the Ph.D. grant of the Fulbright foreign student program.}

\bibliography{Refs}

\begin{thebibliography}{60}%
\makeatletter
\providecommand \@ifxundefined [1]{%
 \@ifx{#1\undefined}
}%
\providecommand \@ifnum [1]{%
 \ifnum #1\expandafter \@firstoftwo
 \else \expandafter \@secondoftwo
 \fi
}%
\providecommand \@ifx [1]{%
 \ifx #1\expandafter \@firstoftwo
 \else \expandafter \@secondoftwo
 \fi
}%
\providecommand \natexlab [1]{#1}%
\providecommand \enquote  [1]{``#1''}%
\providecommand \bibnamefont  [1]{#1}%
\providecommand \bibfnamefont [1]{#1}%
\providecommand \citenamefont [1]{#1}%
\providecommand \href@noop [0]{\@secondoftwo}%
\providecommand \href [0]{\begingroup \@sanitize@url \@href}%
\providecommand \@href[1]{\@@startlink{#1}\@@href}%
\providecommand \@@href[1]{\endgroup#1\@@endlink}%
\providecommand \@sanitize@url [0]{\catcode `\\12\catcode `\$12\catcode
  `\&12\catcode `\#12\catcode `\^12\catcode `\_12\catcode `\%12\relax}%
\providecommand \@@startlink[1]{}%
\providecommand \@@endlink[0]{}%
\providecommand \url  [0]{\begingroup\@sanitize@url \@url }%
\providecommand \@url [1]{\endgroup\@href {#1}{\urlprefix }}%
\providecommand \urlprefix  [0]{URL }%
\providecommand \Eprint [0]{\href }%
\providecommand \doibase [0]{http://dx.doi.org/}%
\providecommand \selectlanguage [0]{\@gobble}%
\providecommand \bibinfo  [0]{\@secondoftwo}%
\providecommand \bibfield  [0]{\@secondoftwo}%
\providecommand \translation [1]{[#1]}%
\providecommand \BibitemOpen [0]{}%
\providecommand \bibitemStop [0]{}%
\providecommand \bibitemNoStop [0]{.\EOS\space}%
\providecommand \EOS [0]{\spacefactor3000\relax}%
\providecommand \BibitemShut  [1]{\csname bibitem#1\endcsname}%
\let\auto@bib@innerbib\@empty
\bibitem [{\citenamefont {{Abbott}}\ \emph {et~al.}(2017)\citenamefont
  {{Abbott}} \emph {et~al.}}]{2017ApJ...848L..12A}%
  \BibitemOpen
  \bibfield  {author} {\bibinfo {author} {\bibfnamefont {B.~P.}\ \bibnamefont
  {{Abbott}}} \emph {et~al.},\ }\href {\doibase 10.3847/2041-8213/aa91c9}
  {\bibfield  {journal} {\bibinfo  {journal} {\apjl}\ }\textbf {\bibinfo
  {volume} {848}},\ \bibinfo {eid} {L12} (\bibinfo {year} {2017})}\BibitemShut
  {NoStop}%
\bibitem [{\citenamefont {Aartsen}\ \emph {et~al.}(2018)\citenamefont {Aartsen}
  \emph {et~al.}}]{ic1709022mm}%
  \BibitemOpen
  \bibfield  {author} {\bibinfo {author} {\bibfnamefont {M.~G.}\ \bibnamefont
  {Aartsen}} \emph {et~al.},\ }\href@noop {} {\bibfield  {journal} {\bibinfo
  {journal} {Science}\ }\textbf {\bibinfo {volume} {361}} (\bibinfo {year}
  {2018})}\BibitemShut {NoStop}%
\bibitem [{\citenamefont {{Marka}}\ \emph {et~al.}(2006)\citenamefont {{Marka}}
  \emph {et~al.}}]{LIGOG060660}%
  \BibitemOpen
  \bibfield  {author} {\bibinfo {author} {\bibfnamefont {Z.}~\bibnamefont
  {{Marka}}} \emph {et~al.},\ }\href@noop {} {\bibfield  {journal} {\bibinfo
  {journal} {LIGO Document G060660}\ } (\bibinfo {year} {2006})}\BibitemShut
  {NoStop}%
\bibitem [{\citenamefont {{Seaman}}\ \emph {et~al.}(2006)\citenamefont
  {{Seaman}} \emph {et~al.}}]{2006ivoa.spec.1101S}%
  \BibitemOpen
  \bibfield  {author} {\bibinfo {author} {\bibfnamefont {R.}~\bibnamefont
  {{Seaman}}} \emph {et~al.},\ }\href@noop {} {\enquote {\bibinfo {title} {{Sky
  Event Reporting Metadata (VOEvent) Version 1.11}},}\ }\bibinfo {howpublished}
  {IVOA Recommendation 1 November 2006} (\bibinfo {year} {2006})\BibitemShut
  {NoStop}%
\bibitem [{\citenamefont {{Aso}}\ \emph {et~al.}(2008)\citenamefont {{Aso}}
  \emph {et~al.}}]{2008CQGra..25k4039A}%
  \BibitemOpen
  \bibfield  {author} {\bibinfo {author} {\bibfnamefont {Y.}~\bibnamefont
  {{Aso}}} \emph {et~al.},\ }\href {\doibase 10.1088/0264-9381/25/11/114039}
  {\bibfield  {journal} {\bibinfo  {journal} {Class. Quantum Grav}\ }\textbf
  {\bibinfo {volume} {25}},\ \bibinfo {eid} {114039} (\bibinfo {year}
  {2008})}\BibitemShut {NoStop}%
\bibitem [{\citenamefont {{Abbott}}\ \emph {et~al.}(2008)\citenamefont
  {{Abbott}} \emph {et~al.}}]{2008CQGra..25k4051A}%
  \BibitemOpen
  \bibfield  {author} {\bibinfo {author} {\bibfnamefont {B.}~\bibnamefont
  {{Abbott}}} \emph {et~al.},\ }\href {\doibase 10.1088/0264-9381/25/11/114051}
  {\bibfield  {journal} {\bibinfo  {journal} {Class. Quantum Grav}\ }\textbf
  {\bibinfo {volume} {25}},\ \bibinfo {eid} {114051} (\bibinfo {year}
  {2008})}\BibitemShut {NoStop}%
\bibitem [{\citenamefont {{van Elewyck}}\ \emph {et~al.}(2009)\citenamefont
  {{van Elewyck}} \emph {et~al.}}]{2009IJMPD..18.1655V}%
  \BibitemOpen
  \bibfield  {author} {\bibinfo {author} {\bibfnamefont {V.}~\bibnamefont {{van
  Elewyck}}} \emph {et~al.},\ }\href {\doibase 10.1142/S0218271809015655}
  {\bibfield  {journal} {\bibinfo  {journal} {Int. J. Mod. Phys. D}\ }\textbf
  {\bibinfo {volume} {18}},\ \bibinfo {pages} {1655} (\bibinfo {year}
  {2009})}\BibitemShut {NoStop}%
\bibitem [{\citenamefont {{M{\'a}rka}}\ \emph {et~al.}(2010)\citenamefont
  {{M{\'a}rka}} \emph {et~al.}}]{2010JPhCS.243a2001M}%
  \BibitemOpen
  \bibfield  {author} {\bibinfo {author} {\bibfnamefont {S.}~\bibnamefont
  {{M{\'a}rka}}} \emph {et~al.},\ }in\ \href {\doibase
  10.1088/1742-6596/243/1/012001} {\emph {\bibinfo {booktitle} {Journal of
  Physics Conference Series}}},\ \bibinfo {series} {Journal of Physics
  Conference Series}, Vol.\ \bibinfo {volume} {243}\ (\bibinfo {year} {2010})\
  p.\ \bibinfo {pages} {012001}\BibitemShut {NoStop}%
\bibitem [{\citenamefont {{Baret}}\ \emph {et~al.}(2011)\citenamefont {{Baret}}
  \emph {et~al.}}]{2011APh....35....1B}%
  \BibitemOpen
  \bibfield  {author} {\bibinfo {author} {\bibfnamefont {B.}~\bibnamefont
  {{Baret}}} \emph {et~al.},\ }\href {\doibase
  10.1016/j.astropartphys.2011.04.001} {\bibfield  {journal} {\bibinfo
  {journal} {Astropart. Phys.}\ }\textbf {\bibinfo {volume} {35}},\ \bibinfo
  {pages} {1} (\bibinfo {year} {2011})}\BibitemShut {NoStop}%
\bibitem [{\citenamefont {{M{\'a}rka}}\ \emph {et~al.}(2011)\citenamefont
  {{M{\'a}rka}} \emph {et~al.}}]{2011CQGra..28k4013M}%
  \BibitemOpen
  \bibfield  {author} {\bibinfo {author} {\bibfnamefont {S.}~\bibnamefont
  {{M{\'a}rka}}} \emph {et~al.},\ }\href {\doibase
  10.1088/0264-9381/28/11/114013} {\bibfield  {journal} {\bibinfo  {journal}
  {Class. Quantum Grav}\ }\textbf {\bibinfo {volume} {28}},\ \bibinfo {eid}
  {114013} (\bibinfo {year} {2011})}\BibitemShut {NoStop}%
\bibitem [{\citenamefont {{Chassande-Mottin}}\ \emph
  {et~al.}(2011)\citenamefont {{Chassande-Mottin}}, \citenamefont {{Hendry}},
  \citenamefont {{Sutton}},\ and\ \citenamefont
  {{M{\'a}rka}}}]{2011GReGr..43..437C}%
  \BibitemOpen
  \bibfield  {author} {\bibinfo {author} {\bibfnamefont {E.}~\bibnamefont
  {{Chassande-Mottin}}}, \bibinfo {author} {\bibfnamefont {M.}~\bibnamefont
  {{Hendry}}}, \bibinfo {author} {\bibfnamefont {P.~J.}\ \bibnamefont
  {{Sutton}}}, \ and\ \bibinfo {author} {\bibfnamefont {S.}~\bibnamefont
  {{M{\'a}rka}}},\ }\href {\doibase 10.1007/s10714-010-1019-z} {\bibfield
  {journal} {\bibinfo  {journal} {Gen. Rel. Gravit}\ }\textbf {\bibinfo
  {volume} {43}},\ \bibinfo {pages} {437} (\bibinfo {year} {2011})}\BibitemShut
  {NoStop}%
\bibitem [{\citenamefont {{Seaman}}\ \emph {et~al.}(2011)\citenamefont
  {{Seaman}} \emph {et~al.}}]{2011ivoa.spec.0711S}%
  \BibitemOpen
  \bibfield  {author} {\bibinfo {author} {\bibfnamefont {R.}~\bibnamefont
  {{Seaman}}} \emph {et~al.},\ }\href@noop {} {\enquote {\bibinfo {title} {{Sky
  Event Reporting Metadata Version 2.0}},}\ }\bibinfo {howpublished} {IVOA
  Recommendation 11 July 2011} (\bibinfo {year} {2011}),\ \Eprint
  {http://arxiv.org/abs/1110.0523} {arXiv:1110.0523 [astro-ph.IM]} \BibitemShut
  {NoStop}%
\bibitem [{\citenamefont {{Bartos}}\ \emph {et~al.}(2011)\citenamefont
  {{Bartos}}, \citenamefont {{Finley}}, \citenamefont {{Corsi}},\ and\
  \citenamefont {{M{\'a}rka}}}]{2011PhRvL.107y1101B}%
  \BibitemOpen
  \bibfield  {author} {\bibinfo {author} {\bibfnamefont {I.}~\bibnamefont
  {{Bartos}}}, \bibinfo {author} {\bibfnamefont {C.}~\bibnamefont {{Finley}}},
  \bibinfo {author} {\bibfnamefont {A.}~\bibnamefont {{Corsi}}}, \ and\
  \bibinfo {author} {\bibfnamefont {S.}~\bibnamefont {{M{\'a}rka}}},\ }\href
  {\doibase 10.1103/PhysRevLett.107.251101} {\bibfield  {journal} {\bibinfo
  {journal} {\prl}\ }\textbf {\bibinfo {volume} {107}},\ \bibinfo {eid}
  {251101} (\bibinfo {year} {2011})}\BibitemShut {NoStop}%
\bibitem [{\citenamefont {{Baret}}\ \emph
  {et~al.}(2012{\natexlab{a}})\citenamefont {{Baret}} \emph
  {et~al.}}]{2012JPhCS.363a2022B}%
  \BibitemOpen
  \bibfield  {author} {\bibinfo {author} {\bibfnamefont {B.}~\bibnamefont
  {{Baret}}} \emph {et~al.},\ }in\ \href {\doibase
  10.1088/1742-6596/363/1/012022} {\emph {\bibinfo {booktitle} {Journal of
  Physics Conference Series}}},\ \bibinfo {series} {Journal of Physics
  Conference Series}, Vol.\ \bibinfo {volume} {363}\ (\bibinfo {year} {2012})\
  p.\ \bibinfo {pages} {012022}\BibitemShut {NoStop}%
\bibitem [{\citenamefont {{Baret}}\ \emph
  {et~al.}(2012{\natexlab{b}})\citenamefont {{Baret}} \emph
  {et~al.}}]{2012PhRvD..85j3004B}%
  \BibitemOpen
  \bibfield  {author} {\bibinfo {author} {\bibfnamefont {B.}~\bibnamefont
  {{Baret}}} \emph {et~al.},\ }\href {\doibase 10.1103/PhysRevD.85.103004}
  {\bibfield  {journal} {\bibinfo  {journal} {\prd}\ }\textbf {\bibinfo
  {volume} {85}},\ \bibinfo {eid} {103004} (\bibinfo {year}
  {2012}{\natexlab{b}})}\BibitemShut {NoStop}%
\bibitem [{\citenamefont {{Bartos}}\ \emph {et~al.}(2012)\citenamefont
  {{Bartos}}, \citenamefont {{Dasgupta}},\ and\ \citenamefont
  {{M{\'a}rka}}}]{2012PhRvD..86h3007B}%
  \BibitemOpen
  \bibfield  {author} {\bibinfo {author} {\bibfnamefont {I.}~\bibnamefont
  {{Bartos}}}, \bibinfo {author} {\bibfnamefont {B.}~\bibnamefont
  {{Dasgupta}}}, \ and\ \bibinfo {author} {\bibfnamefont {S.}~\bibnamefont
  {{M{\'a}rka}}},\ }\href {\doibase 10.1103/PhysRevD.86.083007} {\bibfield
  {journal} {\bibinfo  {journal} {\prd}\ }\textbf {\bibinfo {volume} {86}},\
  \bibinfo {eid} {083007} (\bibinfo {year} {2012})}\BibitemShut {NoStop}%
\bibitem [{\citenamefont {{Smith}}\ \emph {et~al.}(2013)\citenamefont {{Smith}}
  \emph {et~al.}}]{2013APh....45...56S}%
  \BibitemOpen
  \bibfield  {author} {\bibinfo {author} {\bibfnamefont {M.~W.~E.}\
  \bibnamefont {{Smith}}} \emph {et~al.},\ }\href {\doibase
  10.1016/j.astropartphys.2013.03.003} {\bibfield  {journal} {\bibinfo
  {journal} {Astropart. Phys.}\ }\textbf {\bibinfo {volume} {45}},\ \bibinfo
  {pages} {56} (\bibinfo {year} {2013})}\BibitemShut {NoStop}%
\bibitem [{\citenamefont {{Bartos}}\ \emph
  {et~al.}(2013{\natexlab{a}})\citenamefont {{Bartos}}, \citenamefont
  {{Brady}},\ and\ \citenamefont {{M{\'a}rka}}}]{2013CQGra..30l3001B}%
  \BibitemOpen
  \bibfield  {author} {\bibinfo {author} {\bibfnamefont {I.}~\bibnamefont
  {{Bartos}}}, \bibinfo {author} {\bibfnamefont {P.}~\bibnamefont {{Brady}}}, \
  and\ \bibinfo {author} {\bibfnamefont {S.}~\bibnamefont {{M{\'a}rka}}},\
  }\href {\doibase 10.1088/0264-9381/30/12/123001} {\bibfield  {journal}
  {\bibinfo  {journal} {Class. Quantum Grav}\ }\textbf {\bibinfo {volume}
  {30}},\ \bibinfo {eid} {123001} (\bibinfo {year}
  {2013}{\natexlab{a}})}\BibitemShut {NoStop}%
\bibitem [{\citenamefont {{Adri{\'a}n-Mart{\'{\i}}nez}}\ \emph
  {et~al.}(2013)\citenamefont {{Adri{\'a}n-Mart{\'{\i}}nez}} \emph
  {et~al.}}]{2013JCAP...06..008A}%
  \BibitemOpen
  \bibfield  {author} {\bibinfo {author} {\bibfnamefont {S.}~\bibnamefont
  {{Adri{\'a}n-Mart{\'{\i}}nez}}} \emph {et~al.},\ }\href {\doibase
  10.1088/1475-7516/2013/06/008} {\bibfield  {journal} {\bibinfo  {journal}
  {\jcap}\ }\textbf {\bibinfo {volume} {6}},\ \bibinfo {eid} {008} (\bibinfo
  {year} {2013})}\BibitemShut {NoStop}%
\bibitem [{\citenamefont {{Bartos}}\ \emph
  {et~al.}(2013{\natexlab{b}})\citenamefont {{Bartos}}, \citenamefont
  {{Beloborodov}}, \citenamefont {{Hurley}},\ and\ \citenamefont
  {{M{\'a}rka}}}]{2013PhRvL.110x1101B}%
  \BibitemOpen
  \bibfield  {author} {\bibinfo {author} {\bibfnamefont {I.}~\bibnamefont
  {{Bartos}}}, \bibinfo {author} {\bibfnamefont {A.~M.}\ \bibnamefont
  {{Beloborodov}}}, \bibinfo {author} {\bibfnamefont {K.}~\bibnamefont
  {{Hurley}}}, \ and\ \bibinfo {author} {\bibfnamefont {S.}~\bibnamefont
  {{M{\'a}rka}}},\ }\href {\doibase 10.1103/PhysRevLett.110.241101} {\bibfield
  {journal} {\bibinfo  {journal} {\prl}\ }\textbf {\bibinfo {volume} {110}},\
  \bibinfo {eid} {241101} (\bibinfo {year} {2013}{\natexlab{b}})}\BibitemShut
  {NoStop}%
\bibitem [{\citenamefont {{Ando}}\ \emph {et~al.}(2013)\citenamefont {{Ando}}
  \emph {et~al.}}]{2013RvMP...85.1401A}%
  \BibitemOpen
  \bibfield  {author} {\bibinfo {author} {\bibfnamefont {S.}~\bibnamefont
  {{Ando}}} \emph {et~al.},\ }\href {\doibase 10.1103/RevModPhys.85.1401}
  {\bibfield  {journal} {\bibinfo  {journal} {Rev. Mod. Phys}\ }\textbf
  {\bibinfo {volume} {85}},\ \bibinfo {pages} {1401} (\bibinfo {year}
  {2013})}\BibitemShut {NoStop}%
\bibitem [{\citenamefont {{Bartos}}\ and\ \citenamefont
  {{M{\'a}rka}}(2014)}]{2014PhRvD..90j1301B}%
  \BibitemOpen
  \bibfield  {author} {\bibinfo {author} {\bibfnamefont {I.}~\bibnamefont
  {{Bartos}}}\ and\ \bibinfo {author} {\bibfnamefont {S.}~\bibnamefont
  {{M{\'a}rka}}},\ }\href {\doibase 10.1103/PhysRevD.90.101301} {\bibfield
  {journal} {\bibinfo  {journal} {\prd}\ }\textbf {\bibinfo {volume} {90}},\
  \bibinfo {eid} {101301} (\bibinfo {year} {2014})}\BibitemShut {NoStop}%
\bibitem [{\citenamefont {{Aartsen}}\ \emph
  {et~al.}(2014{\natexlab{a}})\citenamefont {{Aartsen}} \emph
  {et~al.}}]{2014PhRvD..90j2002A}%
  \BibitemOpen
  \bibfield  {author} {\bibinfo {author} {\bibfnamefont {M.~G.}\ \bibnamefont
  {{Aartsen}}} \emph {et~al.},\ }\href {\doibase 10.1103/PhysRevD.90.102002}
  {\bibfield  {journal} {\bibinfo  {journal} {\prd}\ }\textbf {\bibinfo
  {volume} {90}},\ \bibinfo {eid} {102002} (\bibinfo {year}
  {2014}{\natexlab{a}})}\BibitemShut {NoStop}%
\bibitem [{\citenamefont {{Bartos}}\ and\ \citenamefont
  {{M{\'a}rka}}(2015)}]{2015PhRvL.115w1101B}%
  \BibitemOpen
  \bibfield  {author} {\bibinfo {author} {\bibfnamefont {I.}~\bibnamefont
  {{Bartos}}}\ and\ \bibinfo {author} {\bibfnamefont {S.}~\bibnamefont
  {{M{\'a}rka}}},\ }\href {\doibase 10.1103/PhysRevLett.115.231101} {\bibfield
  {journal} {\bibinfo  {journal} {\prl}\ }\textbf {\bibinfo {volume} {115}},\
  \bibinfo {eid} {231101} (\bibinfo {year} {2015})}\BibitemShut {NoStop}%
\bibitem [{\citenamefont {{Adri{\'a}n-Mart{\'{\i}}nez}}\ \emph
  {et~al.}(2016{\natexlab{a}})\citenamefont {{Adri{\'a}n-Mart{\'{\i}}nez}}
  \emph {et~al.}}]{2016PhRvD..93l2010A}%
  \BibitemOpen
  \bibfield  {author} {\bibinfo {author} {\bibfnamefont {S.}~\bibnamefont
  {{Adri{\'a}n-Mart{\'{\i}}nez}}} \emph {et~al.},\ }\href {\doibase
  10.1103/PhysRevD.93.122010} {\bibfield  {journal} {\bibinfo  {journal}
  {\prd}\ }\textbf {\bibinfo {volume} {93}},\ \bibinfo {eid} {122010} (\bibinfo
  {year} {2016}{\natexlab{a}})}\BibitemShut {NoStop}%
\bibitem [{\citenamefont {{Albert}}\ \emph
  {et~al.}(2017{\natexlab{a}})\citenamefont {{Albert}} \emph
  {et~al.}}]{2017ApJ...850L..35A}%
  \BibitemOpen
  \bibfield  {author} {\bibinfo {author} {\bibfnamefont {A.}~\bibnamefont
  {{Albert}}} \emph {et~al.},\ }\href {\doibase 10.3847/2041-8213/aa9aed}
  {\bibfield  {journal} {\bibinfo  {journal} {\apjl}\ }\textbf {\bibinfo
  {volume} {850}},\ \bibinfo {eid} {L35} (\bibinfo {year}
  {2017}{\natexlab{a}})}\BibitemShut {NoStop}%
\bibitem [{\citenamefont {{Albert}}\ \emph
  {et~al.}(2017{\natexlab{b}})\citenamefont {{Albert}} \emph
  {et~al.}}]{2017PhRvD..96b2005A}%
  \BibitemOpen
  \bibfield  {author} {\bibinfo {author} {\bibfnamefont {A.}~\bibnamefont
  {{Albert}}} \emph {et~al.},\ }\href {\doibase 10.1103/PhysRevD.96.022005}
  {\bibfield  {journal} {\bibinfo  {journal} {\prd}\ }\textbf {\bibinfo
  {volume} {96}},\ \bibinfo {eid} {022005} (\bibinfo {year}
  {2017}{\natexlab{b}})}\BibitemShut {NoStop}%
\bibitem [{\citenamefont {{Bartos}}\ \emph {et~al.}(2017)\citenamefont
  {{Bartos}}, \citenamefont {{Ahrens}}, \citenamefont {{Finley}},\ and\
  \citenamefont {{M{\'a}rka}}}]{2017PhRvD..96b3003B}%
  \BibitemOpen
  \bibfield  {author} {\bibinfo {author} {\bibfnamefont {I.}~\bibnamefont
  {{Bartos}}}, \bibinfo {author} {\bibfnamefont {M.}~\bibnamefont {{Ahrens}}},
  \bibinfo {author} {\bibfnamefont {C.}~\bibnamefont {{Finley}}}, \ and\
  \bibinfo {author} {\bibfnamefont {S.}~\bibnamefont {{M{\'a}rka}}},\ }\href
  {\doibase 10.1103/PhysRevD.96.023003} {\bibfield  {journal} {\bibinfo
  {journal} {\prd}\ }\textbf {\bibinfo {volume} {96}},\ \bibinfo {eid} {023003}
  (\bibinfo {year} {2017})}\BibitemShut {NoStop}%
\bibitem [{\citenamefont {{Murase}}(2018)}]{kohta-sn-2018}%
  \BibitemOpen
  \bibfield  {author} {\bibinfo {author} {\bibfnamefont {K.}~\bibnamefont
  {{Murase}}},\ }\href {\doibase 10.1103/PhysRevD.97.081301} {\bibfield
  {journal} {\bibinfo  {journal} {\prd}\ }\textbf {\bibinfo {volume} {97}},\
  \bibinfo {eid} {081301} (\bibinfo {year} {2018})}\BibitemShut {NoStop}%
\bibitem [{\citenamefont {{Murase}}\ \emph {et~al.}(2006)\citenamefont
  {{Murase}} \emph {et~al.}}]{kohta2006}%
  \BibitemOpen
  \bibfield  {author} {\bibinfo {author} {\bibfnamefont {K.}~\bibnamefont
  {{Murase}}} \emph {et~al.},\ }\href {\doibase 10.1086/509323} {\bibfield
  {journal} {\bibinfo  {journal} {\apjl}\ }\textbf {\bibinfo {volume} {651}},\
  \bibinfo {pages} {L5} (\bibinfo {year} {2006})}\BibitemShut {NoStop}%
\bibitem [{\citenamefont {{M{\'e}sz{\'a}ros}}(2013)}]{meszaros13}%
  \BibitemOpen
  \bibfield  {author} {\bibinfo {author} {\bibfnamefont {P.}~\bibnamefont
  {{M{\'e}sz{\'a}ros}}},\ }\href {\doibase 10.1016/j.astropartphys.2012.03.009}
  {\bibfield  {journal} {\bibinfo  {journal} {Astropart. Phys.}\ }\textbf
  {\bibinfo {volume} {43}},\ \bibinfo {pages} {134} (\bibinfo {year}
  {2013})}\BibitemShut {NoStop}%
\bibitem [{\citenamefont {{Kimura}}\ \emph {et~al.}(2018)\citenamefont
  {{Kimura}} \emph {et~al.}}]{2018arXiv180511613K}%
  \BibitemOpen
  \bibfield  {author} {\bibinfo {author} {\bibfnamefont {S.~S.}\ \bibnamefont
  {{Kimura}}} \emph {et~al.},\ }\href@noop {} {\bibfield  {journal} {\bibinfo
  {journal} {arXiv:1805.11613}\ } (\bibinfo {year} {2018})}\BibitemShut
  {NoStop}%
\bibitem [{\citenamefont {{Kimura}}\ \emph {et~al.}(2017)\citenamefont
  {{Kimura}} \emph {et~al.}}]{shigeo2017}%
  \BibitemOpen
  \bibfield  {author} {\bibinfo {author} {\bibfnamefont {S.~S.}\ \bibnamefont
  {{Kimura}}} \emph {et~al.},\ }\href {\doibase 10.3847/2041-8213/aa8d14}
  {\bibfield  {journal} {\bibinfo  {journal} {\apjl}\ }\textbf {\bibinfo
  {volume} {848}},\ \bibinfo {eid} {L4} (\bibinfo {year} {2017})}\BibitemShut
  {NoStop}%
\bibitem [{\citenamefont {{Ioka}}\ \emph {et~al.}(2005)\citenamefont {{Ioka}}
  \emph {et~al.}}]{ioka2005}%
  \BibitemOpen
  \bibfield  {author} {\bibinfo {author} {\bibfnamefont {K.}~\bibnamefont
  {{Ioka}}} \emph {et~al.},\ }\href {\doibase 10.1086/466514} {\bibfield
  {journal} {\bibinfo  {journal} {\apj}\ }\textbf {\bibinfo {volume} {633}},\
  \bibinfo {pages} {1013} (\bibinfo {year} {2005})}\BibitemShut {NoStop}%
\bibitem [{\citenamefont {{Murphy}}\ \emph {et~al.}(2013)\citenamefont
  {{Murphy}} \emph {et~al.}}]{murphy2013}%
  \BibitemOpen
  \bibfield  {author} {\bibinfo {author} {\bibfnamefont {D.}~\bibnamefont
  {{Murphy}}} \emph {et~al.},\ }\href {\doibase 10.1103/PhysRevD.87.103008}
  {\bibfield  {journal} {\bibinfo  {journal} {\prd}\ }\textbf {\bibinfo
  {volume} {87}},\ \bibinfo {eid} {103008} (\bibinfo {year}
  {2013})}\BibitemShut {NoStop}%
\bibitem [{\citenamefont {{Aasi}}\ \emph {et~al.}(2015)\citenamefont {{Aasi}}
  \emph {et~al.}}]{aligo2015}%
  \BibitemOpen
  \bibfield  {author} {\bibinfo {author} {\bibfnamefont {J.}~\bibnamefont
  {{Aasi}}} \emph {et~al.},\ }\href {\doibase 10.1088/0264-9381/32/7/074001}
  {\bibfield  {journal} {\bibinfo  {journal} {Class. Quantum Grav.}\ }\textbf
  {\bibinfo {volume} {32}},\ \bibinfo {eid} {074001} (\bibinfo {year}
  {2015})}\BibitemShut {NoStop}%
\bibitem [{\citenamefont {{Acernese}}\ \emph {et~al.}(2015)\citenamefont
  {{Acernese}} \emph {et~al.}}]{avirgo2015}%
  \BibitemOpen
  \bibfield  {author} {\bibinfo {author} {\bibfnamefont {F.}~\bibnamefont
  {{Acernese}}} \emph {et~al.},\ }\href {\doibase
  10.1088/0264-9381/32/2/024001} {\bibfield  {journal} {\bibinfo  {journal}
  {Class. Quantum Grav.}\ }\textbf {\bibinfo {volume} {32}},\ \bibinfo {eid}
  {024001} (\bibinfo {year} {2015})}\BibitemShut {NoStop}%
\bibitem [{\citenamefont {{Aartsen}}\ \emph {et~al.}(2017)\citenamefont
  {{Aartsen}} \emph {et~al.}}]{icecube2017}%
  \BibitemOpen
  \bibfield  {author} {\bibinfo {author} {\bibfnamefont {M.~G.}\ \bibnamefont
  {{Aartsen}}} \emph {et~al.},\ }\href {\doibase
  10.1088/1748-0221/12/03/P03012} {\bibfield  {journal} {\bibinfo  {journal}
  {J. Instrum}\ }\textbf {\bibinfo {volume} {12}},\ \bibinfo {pages} {P03012}
  (\bibinfo {year} {2017})}\BibitemShut {NoStop}%
\bibitem [{\citenamefont {{Ageron}}\ \emph {et~al.}(2011)\citenamefont
  {{Ageron}} \emph {et~al.}}]{2011NIMPA.656...11A}%
  \BibitemOpen
  \bibfield  {author} {\bibinfo {author} {\bibfnamefont {M.}~\bibnamefont
  {{Ageron}}} \emph {et~al.},\ }\href {\doibase 10.1016/j.nima.2011.06.103}
  {\bibfield  {journal} {\bibinfo  {journal} {Nucl. Instrum. Methods Phys. Res.
  A}\ }\textbf {\bibinfo {volume} {656}},\ \bibinfo {pages} {11} (\bibinfo
  {year} {2011})}\BibitemShut {NoStop}%
\bibitem [{\citenamefont {Aab}\ \emph {et~al.}(2015)\citenamefont {Aab} \emph
  {et~al.}}]{2015arXiv150201323T}%
  \BibitemOpen
  \bibfield  {author} {\bibinfo {author} {\bibfnamefont {A.}~\bibnamefont
  {Aab}} \emph {et~al.},\ }\href@noop {} {\bibfield  {journal} {\bibinfo
  {journal} {arXiv:1502.01323}\ } (\bibinfo {year} {2015})}\BibitemShut
  {NoStop}%
\bibitem [{\citenamefont {{Abbott}}(2018)}]{lrr2018}%
  \BibitemOpen
  \bibfield  {author} {\bibinfo {author} {\bibfnamefont {B.~P.~o.}\
  \bibnamefont {{Abbott}}},\ }\href {\doibase 10.1007/s41114-018-0012-9}
  {\bibfield  {journal} {\bibinfo  {journal} {Living Rev. Relativ}\ }\textbf
  {\bibinfo {volume} {21}},\ \bibinfo {eid} {3} (\bibinfo {year}
  {2018})}\BibitemShut {NoStop}%
\bibitem [{\citenamefont {{Aartsen}}\ \emph
  {et~al.}(2014{\natexlab{b}})\citenamefont {{Aartsen}} \emph
  {et~al.}}]{2014arXiv1412.5106I}%
  \BibitemOpen
  \bibfield  {author} {\bibinfo {author} {\bibfnamefont {M.~G.}\ \bibnamefont
  {{Aartsen}}} \emph {et~al.},\ }\href@noop {} {\bibfield  {journal} {\bibinfo
  {journal} {arXiv:1412.5106}\ } (\bibinfo {year}
  {2014}{\natexlab{b}})}\BibitemShut {NoStop}%
\bibitem [{\citenamefont {{Adri{\'a}n-Mart{\'{\i}}nez}}\ \emph
  {et~al.}(2016{\natexlab{b}})\citenamefont {{Adri{\'a}n-Mart{\'{\i}}nez}}
  \emph {et~al.}}]{2016JPhG...43h4001A}%
  \BibitemOpen
  \bibfield  {author} {\bibinfo {author} {\bibfnamefont {S.}~\bibnamefont
  {{Adri{\'a}n-Mart{\'{\i}}nez}}} \emph {et~al.},\ }\href {\doibase
  10.1088/0954-3899/43/8/084001} {\bibfield  {journal} {\bibinfo  {journal} {J.
  Phys. G}\ }\textbf {\bibinfo {volume} {43}},\ \bibinfo {eid} {084001}
  (\bibinfo {year} {2016}{\natexlab{b}})}\BibitemShut {NoStop}%
\bibitem [{\citenamefont {Albert}\ \emph {et~al.}(2019)\citenamefont {Albert}
  \emph {et~al.}}]{Albert_2019}%
  \BibitemOpen
  \bibfield  {author} {\bibinfo {author} {\bibfnamefont {A.}~\bibnamefont
  {Albert}} \emph {et~al.},\ }\href {\doibase 10.3847/1538-4357/aaf21d}
  {\bibfield  {journal} {\bibinfo  {journal} {The Astrophysical Journal}\
  }\textbf {\bibinfo {volume} {870}},\ \bibinfo {pages} {134} (\bibinfo {year}
  {2019})}\BibitemShut {NoStop}%
\bibitem [{\citenamefont {{Veitch}}\ and\ \citenamefont
  {{Vecchio}}(2010)}]{2010PhRvD..81f2003V}%
  \BibitemOpen
  \bibfield  {author} {\bibinfo {author} {\bibfnamefont {J.}~\bibnamefont
  {{Veitch}}}\ and\ \bibinfo {author} {\bibfnamefont {A.}~\bibnamefont
  {{Vecchio}}},\ }\href {\doibase 10.1103/PhysRevD.81.062003} {\bibfield
  {journal} {\bibinfo  {journal} {\prd}\ }\textbf {\bibinfo {volume} {81}},\
  \bibinfo {eid} {062003} (\bibinfo {year} {2010})}\BibitemShut {NoStop}%
\bibitem [{\citenamefont {{Dupuis}}\ and\ \citenamefont
  {{Woan}}(2005)}]{2005PhRvD..72j2002D}%
  \BibitemOpen
  \bibfield  {author} {\bibinfo {author} {\bibfnamefont {R.~J.}\ \bibnamefont
  {{Dupuis}}}\ and\ \bibinfo {author} {\bibfnamefont {G.}~\bibnamefont
  {{Woan}}},\ }\href {\doibase 10.1103/PhysRevD.72.102002} {\bibfield
  {journal} {\bibinfo  {journal} {\prd}\ }\textbf {\bibinfo {volume} {72}},\
  \bibinfo {eid} {102002} (\bibinfo {year} {2005})}\BibitemShut {NoStop}%
\bibitem [{\citenamefont {{Cornish}}\ and\ \citenamefont
  {{Littenberg}}(2015)}]{2015CQGra..32m5012C}%
  \BibitemOpen
  \bibfield  {author} {\bibinfo {author} {\bibfnamefont {N.~J.}\ \bibnamefont
  {{Cornish}}}\ and\ \bibinfo {author} {\bibfnamefont {T.~B.}\ \bibnamefont
  {{Littenberg}}},\ }\href {\doibase 10.1088/0264-9381/32/13/135012} {\bibfield
   {journal} {\bibinfo  {journal} {Class. Quantum Grav}\ }\textbf {\bibinfo
  {volume} {32}},\ \bibinfo {eid} {135012} (\bibinfo {year}
  {2015})}\BibitemShut {NoStop}%
\bibitem [{\citenamefont {{Singer}}\ and\ \citenamefont
  {{Price}}(2016)}]{2016PhRvD..93b4013S}%
  \BibitemOpen
  \bibfield  {author} {\bibinfo {author} {\bibfnamefont {L.~P.}\ \bibnamefont
  {{Singer}}}\ and\ \bibinfo {author} {\bibfnamefont {L.~R.}\ \bibnamefont
  {{Price}}},\ }\href {\doibase 10.1103/PhysRevD.93.024013} {\bibfield
  {journal} {\bibinfo  {journal} {\prd}\ }\textbf {\bibinfo {volume} {93}},\
  \bibinfo {eid} {024013} (\bibinfo {year} {2016})}\BibitemShut {NoStop}%
\bibitem [{\citenamefont {{Budav{\'a}ri}}\ and\ \citenamefont
  {{Szalay}}(2008)}]{2008ApJ...679..301B}%
  \BibitemOpen
  \bibfield  {author} {\bibinfo {author} {\bibfnamefont {T.}~\bibnamefont
  {{Budav{\'a}ri}}}\ and\ \bibinfo {author} {\bibfnamefont {A.~S.}\
  \bibnamefont {{Szalay}}},\ }\href {\doibase 10.1086/587156} {\bibfield
  {journal} {\bibinfo  {journal} {\apj}\ }\textbf {\bibinfo {volume} {679}},\
  \bibinfo {pages} {301} (\bibinfo {year} {2008})}\BibitemShut {NoStop}%
\bibitem [{\citenamefont {{Ashton}}\ \emph {et~al.}(2018)\citenamefont
  {{Ashton}} \emph {et~al.}}]{2018ApJ...860....6A}%
  \BibitemOpen
  \bibfield  {author} {\bibinfo {author} {\bibfnamefont {G.}~\bibnamefont
  {{Ashton}}} \emph {et~al.},\ }\href {\doibase 10.3847/1538-4357/aabfd2}
  {\bibfield  {journal} {\bibinfo  {journal} {\apj}\ }\textbf {\bibinfo
  {volume} {860}},\ \bibinfo {eid} {6} (\bibinfo {year} {2018})}\BibitemShut
  {NoStop}%
\bibitem [{\citenamefont {{Babu}}\ and\ \citenamefont
  {{Feigelson}}(1997)}]{1997scma.conf.....B}%
  \BibitemOpen
  \bibinfo {editor} {\bibfnamefont {G.~J.}\ \bibnamefont {{Babu}}}\ and\
  \bibinfo {editor} {\bibfnamefont {E.~D.}\ \bibnamefont {{Feigelson}}},\
  eds.,\ \href@noop {} {\emph {\bibinfo {title} {Statistical Challenges in
  Modern Astronomy II}}}\ (\bibinfo {year} {1997})\BibitemShut {NoStop}%
\bibitem [{\citenamefont {{Naylor}}\ \emph {et~al.}(2013)\citenamefont
  {{Naylor}}, \citenamefont {{Broos}},\ and\ \citenamefont
  {{Feigelson}}}]{2013ApJS..209...30N}%
  \BibitemOpen
  \bibfield  {author} {\bibinfo {author} {\bibfnamefont {T.}~\bibnamefont
  {{Naylor}}}, \bibinfo {author} {\bibfnamefont {P.~S.}\ \bibnamefont
  {{Broos}}}, \ and\ \bibinfo {author} {\bibfnamefont {E.~D.}\ \bibnamefont
  {{Feigelson}}},\ }\href {\doibase 10.1088/0067-0049/209/2/30} {\bibfield
  {journal} {\bibinfo  {journal} {\apjs}\ }\textbf {\bibinfo {volume} {209}},\
  \bibinfo {eid} {30} (\bibinfo {year} {2013})}\BibitemShut {NoStop}%
\bibitem [{\citenamefont {{Fan}}\ \emph {et~al.}(2014)\citenamefont {{Fan}},
  \citenamefont {{Messenger}},\ and\ \citenamefont
  {{Heng}}}]{2014ApJ...795...43F}%
  \BibitemOpen
  \bibfield  {author} {\bibinfo {author} {\bibfnamefont {X.}~\bibnamefont
  {{Fan}}}, \bibinfo {author} {\bibfnamefont {C.}~\bibnamefont {{Messenger}}},
  \ and\ \bibinfo {author} {\bibfnamefont {I.~S.}\ \bibnamefont {{Heng}}},\
  }\href {\doibase 10.1088/0004-637X/795/1/43} {\bibfield  {journal} {\bibinfo
  {journal} {\apj}\ }\textbf {\bibinfo {volume} {795}},\ \bibinfo {eid} {43}
  (\bibinfo {year} {2014})}\BibitemShut {NoStop}%
\bibitem [{\citenamefont {{Fan}}\ \emph {et~al.}(2017)\citenamefont {{Fan}},
  \citenamefont {{Messenger}},\ and\ \citenamefont
  {{Heng}}}]{2017PhRvL.119r1102F}%
  \BibitemOpen
  \bibfield  {author} {\bibinfo {author} {\bibfnamefont {X.}~\bibnamefont
  {{Fan}}}, \bibinfo {author} {\bibfnamefont {C.}~\bibnamefont {{Messenger}}},
  \ and\ \bibinfo {author} {\bibfnamefont {I.~S.}\ \bibnamefont {{Heng}}},\
  }\href {\doibase 10.1103/PhysRevLett.119.181102} {\bibfield  {journal}
  {\bibinfo  {journal} {\prl}\ }\textbf {\bibinfo {volume} {119}},\ \bibinfo
  {eid} {181102} (\bibinfo {year} {2017})}\BibitemShut {NoStop}%
\bibitem [{\citenamefont {{LIGO and Virgo
  Collaborations}}(2016)}]{PhysRevLett.116.061102}%
  \BibitemOpen
  \bibfield  {author} {\bibinfo {author} {\bibnamefont {{LIGO and Virgo
  Collaborations}}},\ }\href {\doibase 10.1103/PhysRevLett.116.061102}
  {\bibfield  {journal} {\bibinfo  {journal} {Phys. Rev. Lett.}\ }\textbf
  {\bibinfo {volume} {116}},\ \bibinfo {pages} {061102} (\bibinfo {year}
  {2016})}\BibitemShut {NoStop}%
\bibitem [{\citenamefont {{Braun}}\ \emph {et~al.}(2008)\citenamefont {{Braun}}
  \emph {et~al.}}]{2008APh....29..299B}%
  \BibitemOpen
  \bibfield  {author} {\bibinfo {author} {\bibfnamefont {J.}~\bibnamefont
  {{Braun}}} \emph {et~al.},\ }\href {\doibase
  10.1016/j.astropartphys.2008.02.007} {\bibfield  {journal} {\bibinfo
  {journal} {Astropart. Phys.}\ }\textbf {\bibinfo {volume} {29}},\ \bibinfo
  {pages} {299} (\bibinfo {year} {2008})}\BibitemShut {NoStop}%
\bibitem [{\citenamefont {{Aartsen}}\ \emph {et~al.}(2016)\citenamefont
  {{Aartsen}} \emph {et~al.}}]{2016JInst..1111009I}%
  \BibitemOpen
  \bibfield  {author} {\bibinfo {author} {\bibfnamefont {M.~G.}\ \bibnamefont
  {{Aartsen}}} \emph {et~al.},\ }\href {\doibase
  10.1088/1748-0221/11/11/P11009} {\bibfield  {journal} {\bibinfo  {journal}
  {J. Instrum}\ }\textbf {\bibinfo {volume} {11}},\ \bibinfo {pages} {P11009}
  (\bibinfo {year} {2016})}\BibitemShut {NoStop}%
\bibitem [{\citenamefont {{Abadie}}\ \emph {et~al.}(2010)\citenamefont
  {{Abadie}} \emph {et~al.}}]{2010CQGra..27q3001A}%
  \BibitemOpen
  \bibfield  {author} {\bibinfo {author} {\bibfnamefont {J.}~\bibnamefont
  {{Abadie}}} \emph {et~al.},\ }\href {\doibase 10.1088/0264-9381/27/17/173001}
  {\bibfield  {journal} {\bibinfo  {journal} {Class. Quantum Grav}\ }\textbf
  {\bibinfo {volume} {27}},\ \bibinfo {eid} {173001} (\bibinfo {year}
  {2010})}\BibitemShut {NoStop}%
\bibitem [{\citenamefont {{Bartos}}\ \emph {et~al.}(2018)\citenamefont
  {{Bartos}} \emph {et~al.}}]{2018arXiv181011467B}%
  \BibitemOpen
  \bibfield  {author} {\bibinfo {author} {\bibfnamefont {I.}~\bibnamefont
  {{Bartos}}} \emph {et~al.},\ }\href@noop {} {\bibfield  {journal} {\bibinfo
  {journal} {arXiv:1810.11467}\ } (\bibinfo {year} {2018})}\BibitemShut
  {NoStop}%
\bibitem [{\citenamefont {{Waxman}}\ and\ \citenamefont
  {{Bahcall}}(1997)}]{1997PhRvL..78.2292W}%
  \BibitemOpen
  \bibfield  {author} {\bibinfo {author} {\bibfnamefont {E.}~\bibnamefont
  {{Waxman}}}\ and\ \bibinfo {author} {\bibfnamefont {J.}~\bibnamefont
  {{Bahcall}}},\ }\href {\doibase 10.1103/PhysRevLett.78.2292} {\bibfield
  {journal} {\bibinfo  {journal} {\prl}\ }\textbf {\bibinfo {volume} {78}},\
  \bibinfo {pages} {2292} (\bibinfo {year} {1997})}\BibitemShut {NoStop}%
\end{thebibliography}%

\end{document}